\documentclass[twocolumn,english,showpacs,preprintnumbers,amsmath,amssymb,superscriptaddress]{revtex4}
\usepackage[T1]{fontenc}
\usepackage[latin9]{inputenc}
\usepackage{amsmath}
\usepackage{amssymb}
\usepackage{esint}

\makeatletter
\@ifundefined{textcolor}{}
{%
 \definecolor{BLACK}{gray}{0}
 \definecolor{WHITE}{gray}{1}
 \definecolor{RED}{rgb}{1,0,0}
 \definecolor{GREEN}{rgb}{0,1,0}
 \definecolor{BLUE}{rgb}{0,0,1}
 \definecolor{CYAN}{cmyk}{1,0,0,0}
 \definecolor{MAGENTA}{cmyk}{0,1,0,0}
 \definecolor{YELLOW}{cmyk}{0,0,1,0}
}

\@ifundefined{definecolor}{\usepackage{color}}{}
\usepackage{amsfonts}
\usepackage{fancyhdr}\usepackage{fancybox}\usepackage{ulem}

\usepackage{float}

\usepackage{amsthm}

\def\loweq@align#1#2{\lower.6ex\vbox{\baselineskip\z@skip\lineskip\z@
    \ialign{$\m@th#1\hfil##\hfil$\crcr#2\crcr=\crcr}}}
\def\lowsim@align#1#2{\lower.6ex\vbox{\baselineskip\z@skip\lineskip\z@
    \ialign{$\m@th#1\hfil##\hfil$\crcr#2\crcr\sim\crcr}}}
\def\geqq{\mathrel{\mathpalette\loweq@align >}}
\def\leqq{\mathrel{\mathpalette\loweq@align <}}
\def\grsim{\mathrel{\mathpalette\lowsim@align >}}
\def\lesssim{\mathrel{\mathpalette\lowsim@align <}}
\def\gsim{\mathrel{\mathpalette\lowsim@align >}}
\def\lsim{\mathrel{\mathpalette\lowsim@align <}}

\newcommand{\grless}{ {\, \raise-.24em\hbox{$<$} \hspace{-0.8em} \raise.31em\hbox{$>$}\, } }
\newcommand{\lessgr}{ {\, \raise-.24em\hbox{$>$} \hspace{-0.8em} \raise.31em\hbox{$<$}\, } }
\newfont{\bg}{cmr10 scaled\magstep4}                    
\newcommand{\bigzerou}{\smash{\lower1.7ex\hbox{\bg 0}}}

\newcommand{\crl}[1]{[-\infty,\infty]}

\usepackage{babel}

\makeatother

\usepackage{babel}
\begin{document}
\pagecolor{white}

\title{Time Evolution Operators for Periodic SU(3)}

\author{Peter G. Morrison}

\email{nanoscope@outlook.com}

\selectlanguage{english}%

\affiliation{Environmental Control, Sydney, Australia}

\date{July 17, 2019}
\begin{abstract}
\noindent We present an outline of a technique to associate certain
methods from time optimal quantum control with various transforms
on SU(3). Unitary operators are taken from certain time dependent
Hamiltonians and transformation laws are derived. A methodological
framework for development of solution for these types of problems
with a non-simple geometric framework is presented that is generally
applicable. We give an expansive overview of the field of SU(3), its
meaning, purpose and context within the presented results. Exact solutions
for periodic curves on SU(3) are provided, along with the operators
and physical framework that surrounds them. A comprehensive review
of the literature and established results is given for reference.
Discussion is given in full to effective approaches that may be given
both to the teaching of the subject, and the development of further
results within the science. We conclude the paper with an overview
of new avenues of investigation that are opened through the results
that are presented.


\end{abstract}


\maketitle

\section{Introduction}

It is possible to find unitary operators for time dependent Hamiltonian
matrices, using the technique of time optimal state control. In this
paper we shall consider the unitary transformations that result from
consideration of particular Hamiltonians on the space of qutrits.
In doing so, we aim to objectively analyse the geometry of SU(3),
both as an exercise in understanding of this space, which is related
to ace \cite{Zweig} geometry and fundamental particles via the standard
model, and also to further the development of qutrit computation.
This space contains essential degeneracies that we shall give some
results for. These degeneracies stem from the existence of a zero-eigenstate,
and it requires that we modify our approach to deal with this situation.
A rich matrix calculus exists on this space of qutrits, however, the
situation is complicated by the many-valued-ness in decompositions
one may use to produce Hamiltonian matrices. This existence of multiple
ways to generate optimal paths on different subspaces within a quantum
dynamical system shall be analysed in great detail for this paper. 

The analysis of the special unitary group has a long history, both
within physics and mathematics. However, due to the simplicity of
solutions in even-numbered dimensions, and the obvious physical applications,
the area of higher order symmetry is often left aside. We shall show
that quite basic assumptions, and a very small number of dynamic equations
can result in the development of a great deal of understanding. In
doing so, we shall address the symmetry groups related to certain
dynamic systems, and at the same time learn about how the notion of
a periodic state extends to higher dimensional objects, resulting
in measurable consequences for the laboratory.

We shall give a compendium of the results from different groups that
have explored SU(3) and the implications from a theoretical and experimental
perspective of our results. The relative scarcity of work that has
been done in this area makes it of value to list what literature exists
on the topic, and which research is engaged and likely to be of potential
gain in the future in light of these new results. SU(3) is an area
of rich geometry and calculus, but its difficult nature has previously
made it relatively inaccessible. Nonetheless, much productive work
has been carried out in the fields of projective geometry, differential
Lie groups, computer experimentation, and nuclear shell models.

It is important to note the implications of this work as well; by
providing new, exact solutions and a unique set of matrices to describe
this group, we imply both experimental and theoretical results. The
link between Lie groups of matrices and differential operators is
a well trodden path. The link to symmetry of physical systems, conserved
quantities and quantisation is the foundation of modern physics. However,
due to the difficulty of solving SU(3) invariant differential equations,
much of the work in theoretical terms has been in modelling and computer
simulation to understand the physical behaviour of exotic quantum
states which obey this symmetry. This is of value in understanding
the nature of the physical objects we are dealing within this paper,
and for this reason we shall cover this in depth. From an experimental
perspective, we shall cover the topics of quark \cite{Gellmann} confinement
and measurements of excited states of heavy-nuclei atoms \cite{Neeman}.
This will enable us to cover an extensive analysis of the topic of
quadrupole-quadrupole interaction, its ubiquitous nature, and the
results that have been found from experiments that have been carried
out in particle colliders and the like. And so, onwards. 

The paper will proceed as follows; firstly, we shall examine the technicalities
in communication of the subject, and take a comprehensive overview
of the state of the science. We will look at the different groups
that are examining this sort of dynamical symmetry, across a range
of different disciplines within mathematics and physics from an experimental
and theoretical perspective. We will then take account of the terms
to be used in the calculation, and show how one may go about developing
the problem as it stands. Time evolution operators will then be calculated
from one perspective; taking into account the laws of motion and symmetry
of the solution, we will then pivot and calculate from another perspective
to ensure the consistency of our method. This having been achieved,
points of discussion and generalisation will be proposed, thereby
giving \textit{pars una imperium in medio chao.}

\section{The Essential Difficulty}

The topic of time dependent quantum mechanics is difficult to teach,
even to experts, maybe more so. Since it is the exclusive domain of
our research let us briefly examine some reasons why this is the case
using some recent work that has been carried out in \cite{Johnston,Marshman,Singh,Singh-1}.
This work has involved the development of collaborative survey and
testing techniques across a number of universities for both introductory
and high level quantum mechanics students in order to understand the
learning process. Let us briefly recap their results from \cite{Singh-1};
we shall then offer some remedies, both from the outcomes in \cite{Singh}
as well as some insight and experience. The principal reasons they
give, across measured trials, for misunderstanding time dependent
quantum mechanics are as follows; the incorrect belief that the time
independent Schrödinger equation is the most fundamental equation
in quantum mechanics; inability to distinguish between $e^{-i\hat{H}t}$
and $e^{-i\int\hat{H}ds}$; incorrect belief that the time evolution
of a wave function is always via an overall phase factor; the incorrect
belief that for a time-independent Hamiltonian, the wave function
does not depend on time. These are all cardinal sins in the world
of time dependent quantum mechanics. It is also important to consider
the other major factors measured in this study for quantum mechanics
in general. They list other associated learning issues that include
the incorrect belief that $\hat{H}\Psi=E\Psi$ for any possible wavefunction
$\Psi$, difficulties with mathematical representations of non-stationary
wavefunctions, and the problems associated with the diverse representations
of a wavefunction. We note that this study has not just been applied
at the undergraduate level, but to upper-level quantum mechanics students.
One can expect a similar result from many experts on the basis of
linear extrapolation once they are moved out of their zone of intuition.
Johnston et al. in \cite{Johnston} states that ``...students have
constructed mental models to conceptualize (sic) the abstract concepts
of quantum mechanics which have little support from anything else
in their experience'' and furthermore that ``...when asked directly
why quantum mechanics is difficult most students answer something
to the effect: 'It's all mathematics'''. 

These symptoms all have a common cause. Let us consider what it might
be, and how best to address it. The problems stem from the transition
stage; one goes through the process of learning classical mechanics,
and is then faced with the transition to something completely different.
That is difficult in and of itself. What we propose is furthermore
complicated, as it is a further transition past the initial jump.
One can immediately appreciate the difficulties that the mind is forced
to go through to make this hurdle.

The leap is made by making it earlier, harder, faster and more explorative
of the concept, both by students and teachers. By postponing the introduction
of these fundamental physical concepts, we actually retard development.
On an expert level, we must strive to introduce the complicated subjects
first and then exhibit the special cases as the unique examples that
they are. It serves no good purpose to have students struggling with
concatenated levels of difficulty, trying to build an intuition from
one concept only to have it fail in another; dealing with the difficulty
earlier will stem the flow.

With respect to our particular domain, time dependent quantum mechanics,
we must stress the following. Time dependent quantum mechanics is
the primary evolutionary factor in every quantum system without exception.
Time independent quantum mechanics is and always will be a special
case of the former and not vice versa. The intuitions developed within
systems that do not change with time will not ever work here. For
that reason it is important to approach it with fresh eyes, open to
the concept. As our intuition will fail here, all we have resort to
are the tools of mathematics, and in particular, matrices. These operators
will represent the time ordering of our system and preserve all the
relations between the fundamental components of the space we will
engage with, SU(3). We cannot avoid the mathematics, so we must in
fact swim against the tide and embrace it as our only chance of understanding
the complex scenarios we shall see.

\section{Review}

\subsection{Nuclear Models, Heavy Nuclei and Quadrupole Interactions}

We shall first cover the results that are able to be compared from
the groups that are working on nuclear models, heavy nuclei and quadrupole
interactions . These groups share several major focal points; they
are generally examining the effects on the quantum states within nuclear
drops due to spherical deformation using spherical polynomial expansions.
This deformation can take the form of hard inelastic scattering, where
the incident particle interacts deeply with the nucleus before recoiling.
The nucleus also recoils, and the compounding effects cause rearrangement
in the internal wavefunctions which are used to model the internal
atomic states. We note the strong overlap between theoretical teams
and experimental results which is present in this area.

For a historical overview of the development of the SU(3) model in
nuclear physics, consult \cite{Arima} for an overview of the development
of the nuclear quadrupole interaction model. \cite{Almeida} c. 2004
contains a good reference for physical implementation of SU(3) states,
and discussion of the topic from the angle of nuclear quadrupole moments.
Much work has been done from the continuous wavefunction perspective
using the original results of Elliot (1958-1963) \cite{Elliot}, who
developed an SU(3) formalism to deal with quadrupole-quadrupole coupling
elements in a Hamiltonian schemata. These groups are focused mostly
on the implementation of modelling using continuous state formalism,
and we shall not use this technique here. Instead of using sets of
orthogonal polynomials to represent the state, we shall encode our
dynamics and symmetry directly in the matrices themselves and avoid
the differential algebra entirely. Doing so has interesting consequences,
as we shall see. In \cite{Almeida}, the authors use an expansion
of the angular momentum state to model the quadrupole-quadrupole interactions
in Be isotopes. It is a good reference for some heavy nuclei calculations
that may be compared with experiment. \cite{Dytrych} contains an
overview of collective modes in light nuclei, examining the SU(3)
coupling modes for lithium, beryllium and helium. They show that these
states have strong SU(3) deformations from nuclear quadrupole interactions,
which leads to complementary low values in spin associated with their
nuclear model. For calculation of collective motion in Kr, \cite{Petrovici}
performed earlier work that provides a point of comparison with currently
available data, using an interesting triangular configuration of states
to simulate heavy nuclear motion.

For a generalised reference, \cite{Harvey} provides an in-depth analysis
of the development of the quadrupole interaction model. In \cite{Marumori},
the authors consider a self-consistent field method for modelling
nuclear collective motion. They examine in passing the topic that
we shall focus on, being the multiple optimal paths that exist on
this space. There are known parallels between the calculations using
matrices to represent non-commutativity and that of composition of
differential Lie groups which means that the translation of the results
we shall present may have interesting consequences. For more complex
models, \cite{Otsuka} looked at at interacting boson modes in order
to model collective states of quadrupole-quadrupole collective motion.
Finally, \cite{Rosensteel} contains an overview of a computation
relating to the collective boson modes in neon. This also uses a quadrupole
expansion over spherical polynomials to calculate physical properties
of a deformed nuclear droplet.

\subsection{Bose-Einstein Condensates/Coherent States}

The collective motion phenomenon has also been probed in the BEC regime,
although not to the same extent as with heavy nuclei. The use of these
types of physical systems and experimental investigations is important
with respect to our results, as it provides a well-known, proven testing
ground for understanding the sorts of phenomena we shall describe.
\cite{Alodjants} describe methods for creating SU(3) polarisation
states in BECs using a coherent state formalism and useful mathematical
physics that could be applied to understand the system that we shall
be examining. For the SU(3) orbital Kondo effect, consult \cite{Nishida}
for a conductance expansion that demonstrates SU(3) effects from ultracold
atoms using a second quantisation schemata. \cite{Tam} address SU(3)
quantum interferometry with photon pulses; using a comparatively simplistic
method relative to the techniques we shall describe. For further mathematical
research on SU(3) coherent states, \cite{Chaturvedi} examined the
multiplicity problem which we shall encounter using a tensor analytic
approach. Finally, \cite{Mathur} is a concisive reference for much
work that has been done on coherent states and contains a solid mathematical
description of the apparatus that lies behind the curtain of SU(3)
coherent state methodology. Topics examined that are of particular
interest are the way in which they deal with path integration on this
type of complex space with degeneracy. Many of the transformations
they cover are of validity in the regime we shall be exploring. Although
we shall not need them, Wigner coefficients and their relationship
to SU(3) generators are important in the differential operator schemata,
and the relationships between singlet states on SU(3) and the relationships
with the coherent states may be found in \cite{Anishetty}.

\subsection{Simulation of SU(3) in Quantum Systems}

We now proceed to the topic of computer simulation of SU(3) effects.
This is an important area of current and ongoing investigation and
is useful in its development of understanding of this strange geometry
which defines the quantum states. \cite{Chen} examine a spin-2 chain
with emergent SU(3) symmetry for ultracold gases. Interestingly, they
concluded that some gases which should only display SU(2) behaviour
at the thermodynamic limit showed SU(3) properties spontaneously.
This is another important point of comparison, both with the computer
simulation and the nominal output of any experiments. \cite{Han}
considered a number of different vortex configurations for a BEC system
similar to those discussed in the prior section. The Kondo effect
with SU(3) properties in spinless triple quantum dots was investigated
in \cite{Lopez}. For more mathematical descriptions, \cite{Clark}
looked at the implementation of a valence bond formalism in a quantum
computation setting, whereas \cite{Evans} contains an outline of
simulation that demonstrates that in SU(3) systems, double phase BECs
can exist.

The prohibitive complexity of SU(3) systems has necessitated the implementation
of these types of computer experiments. That does not, however, negate
their validity in light of increased understanding that we shall provide
through the use of matrix calculus. Instead, it will allow a valuable
point of comparison in the future, that we can use to measure the
difference between techniques of matrix mechanics and wave mechanics,
which may not have the elegant solutions we shall show.

\subsection{Graph Theory}

Although we shall not be using the methods of graph theory, it is
a relevant place in which to observe the relative complexity that
introduction of an essential degeneracy into what was a previously
manageable situation can wreak. For example, in \cite{Chbili}, also
\cite{Ohtsuki}, the authors look at various SU(3) invariants. However,
the calculation is quickly subsumed by difficulties surrounding the
essential non-removable singularity in the punctured manifold. The
classification of knots within \cite{Chbili} and the links to the
invariants in \cite{Ohtsuki} should be explored further in light
of the new results we shall present. The matrix calculus and pattern
of isometric symmetry that shall be demonstrated implies the existence
of certain topological theories, invariants and untwisting congruences
that will have consequences within mathematics. Further explorations
of graph theoretical notions with more reference to mathematical physics
may be found in earlier works in \cite{Mukherjee}, which examined
the application of SU(3) symmetry to calculate ionisation potentials
and Pi-Pi bond energies in transbutadiene using a continuous Hamiltonian
with nuclear spin effects.

\subsection{Quasi-Dynamical Symmetry}

We will now focus on the groups that are working within quasi-dynamical
symmetry. This property is a collective mechanism by which atomic
states and nucleons share rotational motion when in a highly excited
state. A quasi-dynamical symmetry can be seen in some ways as equivalent
to a Hermitian operator which induces a unitary transformation, combined
with a mirror inversion. For reference to works that consider SU(3)
quasi-symmetry that also have aspects considering nuclear motion as
discussed earlier, \cite{Bahri} goes into some depth into how SU(3)
operators are constructed from a continuous state viewpoint using
spherical polynomials. In \cite{Gnutzmann}, the authors go into great
detail into how SU(3) can be used to induce quantum chaos. A good
reference to Casimir invariants and the transition from classical
to quantum chaos is covered with relation to the nuclear shell model
and coherent states. Further works are contained within \cite{Rochford},
which again uses an expansion in spherical polynomials to model the
quadrupole-quadrupole effect. Finally, in \cite{Macek}, interacting
bosonic nuclei effects are calculated that reveal some effects that
will take an increased understanding and importance with regards to
our calculation.

\subsection{SU(3) Geometric Methods}

Geometric methods for SU(3) take a number of forms. Various groups
are working on techniques that range from the use of Lie-theoretic
expansions and geometric algebra, to trigonometric investigations,
geometric phases and sphere-like constructions. We shall discuss these
topics briefly as we shall have indirect recourse to many of the different
branches of geometric analysis that this area has developed to deal
with problems on SU(3).

Firstly, the works of Byrd \cite{Byrd 1,Byrd 2} give a thorough and
comprehensive overview of the entirety of the geometric algebra associated
to SU(3). They give a work-through of what can be achieved using the
exterior algebraic method, and demonstrate many results that are difficult
to show by other techniques. In particular, this method has the advantage
of being relatively independent of the coordinates, and the formulae
may be safely applied to obtain the results. The essential difficulties
of SU(3) when seen from a Lie differential perspective are apparent;
the complexities of the inter-nested partial derivatives make this
a lengthy exposition. The wedge product formalism produces results
that will be complementary to what we produce in this paper; however,
having built all the hardware into our matrices, the hard work for
us shall be mostly done by the matrix multiplication.

For more examinations of the laws of trigonometry associated to the
SU(3) space, consult \cite{Aslaksen} for an earlier account of the
various ways in which trigonometric expansions of functions can be
achieved using trace polynomials and spherical geometry. \cite{Khanna}
is an earlier reference which contains some useful results that can
form a point of comparison with the results of \cite{Byrd 1,Byrd 2};
this work is a good place to develop a feel for the geometric nature
of three state quantum systems and how they can be approached. The
results of \cite{Mallesh} expand upon the initial methods from \cite{Khanna}
and result in some interesting geometric phase formulae for this complex
space. These features are generally for adiatic systems, however,
the picture of geometric phase and its physical consideration is a
worthy place to approach the types of time dependent systems we shall
consider in this paper.

For further expositions on the geometry of SU(3) with particular application
to qutrits, \cite{Li} gives an outline of a technique that relies
on a different metric configuration to this paper. Finally, the work
of \cite{Vinjamampathy} provides a good visual reference for some
ways in which these types of complex multidimensional spaces can be
understood in graphical terms by splitting co-ordinates over Bloch
spheres and projections. The results used in their paper are very
similar in flavour to the tasks we shall approach and solve. Related
topics may be found in the notes of \cite{Eade} for vector-matrix
calculus on implementing rotations in multiple dimensions using similar
exponential expansions to those we shall be using in our investigation.

\section{Terminology}

The mathematics we shall be using for this topic is involved, complicated
by the diversity of variables within this matrix space. We shall briefly
discuss the terms and mathematical apparatus used. To begin with,
when referring to SU(3), we refer to the set of unitary operators
$\hat{U}$ that operate on a vector space in three complex dimensions.
A vector will be of the form:
\begin{equation}
\left|\psi\right\rangle =\left[\begin{array}{c}
c_{1}(t)\\
c_{2}(t)\\
c_{3}(t)
\end{array}\right]
\end{equation}
where $c_{j}(t)$ are complex functions of the time, which might be
constant, and $\left\langle \psi\right|=\left(\left|\psi\right\rangle \right)^{\dagger}=\left[\begin{array}{ccc}
c_{1}^{\star}(t), & c_{2}^{\star}(t), & c_{3}^{\star}(t)\end{array}\right]$ is the adjoint or Hermitian conjugate that describes the inner product
relations. The $\hat{U}$'s then take the form of matrix transformations
that map input states to output states in a way that preserves the
inner product which describes the space. This allows us to freely
move from one reference frame to another, safe in the knowledge that
the essential physics is unchanged under the transformation when applied
correctly. These inner product relations are written as $\left\langle \psi\right|\left.\psi\right\rangle =\sum\left|c_{j}(t)\right|^{2}$
in keeping with the original spirit of Dirac \cite{Dirac}.

With regards to operators, which in this space take the form of matrices,
a matrix which has zero along the sum of the diagonals will be denoted
as $\tilde{A}$. Generally, a unitary operator or operator which has
some utility or special significance but not this property will be
denoted $\hat{A}$. For expendable variables that do not form a part
of the calculation, as in the preceding sentences, we shall use the
letter $A$ wherever possible. Matrices that have a point-in-time
dependence, and that do not have an initial and final condition built
in, will be written as $\hat{A}(t)$. A matrix with two times, i.e.
a start and a finish, will be denoted as $\hat{A}(t,s)$ for some
parameters $t,s$ or $\hat{A}(t,t_{0})$ if $t_{0}$ has special significance.
Our calculation will be presented for a particular concrete representation,
so we will not have to resort to the theory of groups, although it
shall be lying behind everything we do. Finally, since we shall have
much use for it, the sum of diagonals of a matrix shall be called
trace; we shall denote the trace of a matrix $\hat{A}$ by $\textrm{Tr}[\hat{A}]$.

\section{Outline of the Problem}

We shall show how one can develop a series of time optimal quantum
control problems which will serve as a particular place for application
of these techniques. We note the existence of an action principle,
akin to Fermat's principle for quantum systems, that optimises the
time taken for state-to-state transfer, over the complex projective
space \cite{Carlini 1,Carlini 2}. Recent advances have extended this
approach to more exotic systems such as mixed states \cite{Carlini 3}
and coupled Ising chains \cite{Carlini 4}. We shall be applying this
method the group defined as SU(3). The following equations will demonstrate
the principles of application; we shall not delve deeply into the
proof.

The steps of the calculation are as follows. We must first apply constraints
to the system. The constraints will take the form of allowing some
transitions while are others to be forbidden. A natural constraint
we shall have is the net overhead energy of the internal states. This
will be represented by an isotropic condition that will hold the vector
that describes the net total energy variance to be some finite length.
The conservation laws are packed into the constraints; this is a key
point which we shall examine in this paper. The remaining procedures
to be followed shall then consist of setting up a Hamiltonian and
constraint matrix full of arbitrary parameters and appropriate symmetry
to describe the physical situation at hand. We shall then calculate
the time dependence of these arbitrary parameters and thereby resolve
the time dependence of the Hamiltonian matrix, which drives the state
transitions within the quantum system. Once this is achieved, we will
use this Hamiltonian matrix to calculate a certain unitary operator
in a diagonal frame of reference using a time dependent transformation.
We will then invert the time dependent transformation using a form
of isometry to solve for the time dependence of the quantum state.
Once we have this, we can calculate all properties of the quantum
system.

For our interests, it is sufficient to note that the procedure relies
on the observation that one may form a Hermitian matrix composed of
a linear combination of all generators that are not contained in the
Hamiltonian. By definition, we shall have $\textrm{Tr}(\tilde{H}\tilde{F})=0$.
We shall term this matrix $\tilde{F}$ to be 'constraint'; the other
piece of the dynamics is to constrain the Hamiltonian to be isotropic,
and of finite total energy, which is achieved by $\textrm{Tr}[\tilde{H}^{2}/2]=k$
for some constant. We note that the application of the Heisenberg
equation, which follows naturally from the formal calculation of the
quantum brachistochrone, allows us to state the following for the
evolution of certain Hermitian operators:
\begin{equation}
i\dfrac{d\hat{A}}{dt}=\tilde{H}\hat{A}-\hat{A}\tilde{H}
\end{equation}
Now, consider the matrix which we define by the equation $\hat{A}=\tilde{H}+\tilde{F}$.
Both the Hamiltonian and constraint are Hermitian matrices of functions
of time, they have all the regular properties one might need from
reasonably behaved operators. In particular, their sum is also Hermitian,
so we may substitute this into the above equation to obtain the following:
\begin{equation}
i\dfrac{d}{dt}\left(\tilde{H}+\tilde{F}\right)=\tilde{H}\tilde{F}-\tilde{F}\tilde{H}
\end{equation}
This is now a series of coupled ordinary differential equations in
the constraint and Hamiltonian, that has to be solved for the time
dependence of the functions that make up the entries in both the Hamiltonian
and constraint. This is then used this to compute an operator which
describes the unitary evolution in time of the system. We may now
note a particular instance of this problem on SU(3), originally considered
in \cite{Morrison 1}. We take the following Hamiltonian and constraint
set: 
\begin{equation}
\tilde{H}(t)=\left[\begin{array}{ccc}
0 & \varepsilon_{1}(t) & 0\\
\overline{\varepsilon}_{1}(t) & 0 & \varepsilon_{2}(t)\\
0 & \overline{\varepsilon}_{2}(t) & 0
\end{array}\right]
\end{equation}

\begin{equation}
\tilde{F}(t)=\left[\begin{array}{ccc}
\omega_{1} & 0 & \kappa\\
0 & -(\omega_{1}+\omega_{2}) & 0\\
\overline{\kappa} & 0 & \omega_{2}
\end{array}\right]
\end{equation}
 where we use the bar to indicate a complex conjugate. This approach
is a modified version of the techniques applied within reference \cite{Morrison 1,Morrison 2,Morrison 3}.
We now begin to apply the mechanics to solve the quantum brachistochrone
equation. Calculating eq. (2) using the matrices above, we obtain
the following differential equations:

\begin{equation}
i\dfrac{d}{dt}\left[\begin{array}{c}
\varepsilon_{1}\\
\overline{\varepsilon}_{1}\\
\varepsilon_{2}\\
\overline{\varepsilon}_{2}
\end{array}\right]=\left[\begin{array}{cccc}
-\omega_{+} & 0 & 0 & -\kappa\\
0 & \omega_{+} & \overline{\kappa} & 0\\
0 & \kappa & -\omega_{-} & 0\\
-\overline{\kappa} & 0 & 0 & \omega_{-}
\end{array}\right]\left[\begin{array}{c}
\varepsilon_{1}\\
\overline{\varepsilon}_{1}\\
\varepsilon_{2}\\
\overline{\varepsilon}_{2}
\end{array}\right]
\end{equation}
as well as $\dot{\omega}_{1}=\dot{\omega}_{2}=0$, $\dot{\kappa}=\dot{\overline{\kappa}}=0$.
Now, for boundary conditions, consider the following operator $\hat{G}=\left[\tilde{H}+\tilde{F}\right]-\textrm{Tr}\left(\hat{P}\left[\tilde{H}+\tilde{F}\right]\right)\hat{P}$.
It is possible to show, using the argument contained in \cite{Carlini 1,Carlini 2}
that this operator evolves according to the Heisenberg equation of
motion and is a proper quantum variable. We shall not be considering
the technicalities involved with mixed states here, rather, we just
require some simple guidance as to how the variables, assumed constant,
are at the initial point of the time evolution. Constructing the operator
as above, and for the proposed state $\left|\psi(0)\right\rangle =\left[\begin{array}{ccc}
1, & 0, & 0\end{array}\right]^{T}$, we may write: 
\begin{equation}
\hat{G}(0)=\left[\begin{array}{ccc}
0 & \varepsilon_{1}(0) & \kappa\\
\overline{\varepsilon}_{1}(0) & -(\omega_{1}+\omega_{2}) & \varepsilon_{2}(0)\\
\overline{\kappa} & \overline{\varepsilon}_{2}(0) & \omega_{2}
\end{array}\right]
\end{equation}
We have the expression for the boundary conditions on the state the
expression $\hat{G}(t)=\left\{ \hat{G}(t),\hat{P}(t)\right\} $ and
hence we conclude that $\omega_{1}=\omega_{2}=0$ by computing the
matrix multiplications and comparing both sides, also that $\varepsilon_{2}(0)=\overline{\varepsilon}_{2}(0)$.
The diagonal elements of the constraint matrix are related to certain
unitary transformations as we shall demonstrate later in this paper.
Note that this choice of boundary condition was not unique and indeed
one can show that, by permutation over different initial states that
taking the diagonal elements of the constraint as zero is independent
of boundary conditions for any of the extremal points. At this juncture,
one could, as in \cite{Morrison 1}, propose a hypothetical terminal
boundary condition and determine the control fields in time, but we
shall see that this is indeed not necessary. As the geometry the state
is evolving through is not simple to picture geometrically, it may
be that some states are reachable, and others not. Indeed, what is
worse, and turns out to be the case for this geometry, is when there
are multiple paths through different subspaces that are mutually indistinguishable.
For this reason, we shall only tentatively postulate the initial condition
to determine the nature of the constraint. The picture is considerably
simplified, one may write the control equations for the Hamiltonian
fields eqn.(3) in the form:
\begin{equation}
i\dfrac{d\xi}{dt}=\hat{\varUpsilon}\xi
\end{equation}
\begin{equation}
\hat{\varUpsilon}=\left[\begin{array}{cccc}
0 & 0 & 0 & -\kappa\\
0 & 0 & \overline{\kappa} & 0\\
0 & \kappa & 0 & 0\\
-\overline{\kappa} & 0 & 0 & 0
\end{array}\right]
\end{equation}
with $\hat{\varUpsilon}^{2}=\left|\kappa\right|^{2}\mathbf{1}$. Computing
the matrix exponential, we find the expression $\exp\left(-it\hat{\varUpsilon}\right)=\mathbf{1}\cos(kt)-\dfrac{i\hat{\varUpsilon}}{k}\sin(kt)$
where $k=\left|\kappa\right|$. We may write the solution for the
control fields as functions of time $\xi(t)=\exp\left(-it\hat{\varUpsilon}\right)\xi(0)$,
and evaluate our solution using the matrix below:
\begin{eqnarray*}
\exp\left(-it\hat{\varUpsilon}\right)=
\end{eqnarray*}

\begin{equation}
\left[\begin{array}{cccc}
\cos kt & 0 & 0 & -e^{-i\theta}\sin kt\\
0 & \cos kt & e^{i\theta}\sin kt & 0\\
0 & e^{-i\theta}\sin kt & \cos kt & 0\\
-e^{i\theta}\sin kt & 0 & 0 & \cos kt
\end{array}\right]
\end{equation}
where $\theta$ is the phase of the constraint. Evaluating this in
full, we obtain the following series of equations for the time evolution
of the Hamiltonian control fields: 
\begin{equation}
\begin{array}{c}
\varepsilon_{1}(t)=\varepsilon_{1}(0)\cos kt\\
\overline{\varepsilon}_{1}(t)=\overline{\varepsilon}_{1}(0)\cos kt\\
\varepsilon_{2}(t)=\overline{\varepsilon}_{1}(0)e^{-i\theta}\sin kt\\
\overline{\varepsilon}_{2}(t)=-\varepsilon_{1}(0)e^{i\theta}\sin kt
\end{array}
\end{equation}
By substitution of the initial time, one immediately writes off the
boundary conditions on the Hamiltonian as $\varepsilon_{2}(0)=\overline{\varepsilon}_{2}(0)=0$.
We must therefore have the following solutions: 
\begin{equation}
\begin{array}{c}
\varepsilon_{1}(t)=R\cos kt\\
\varepsilon_{2}(t)=-iRe^{-i\theta}\sin kt
\end{array}
\end{equation}
which satisfies the expression $\left|\varepsilon_{1}(t)\right|^{2}+\left|\varepsilon_{2}(t)\right|^{2}=R^{2}$,
coming from the isotropic condition of energy $\textrm{Tr}\left[\tilde{H}^{2}/2\right]=R^{2}$.
There is a second global phase but we shall not consider it further
in this analysis as it does not affect the dynamics. Writing down
the Hamiltonian, we obtain: 
\begin{equation}
\tilde{H}(t)=R\left[\begin{array}{ccc}
0 & \cos kt & 0\\
\cos kt & 0 & -ie^{-i\theta}\sin kt\\
0 & ie^{i\theta}\sin kt & 0
\end{array}\right]
\end{equation}
and for the constraint:
\begin{equation}
\tilde{F}(t)=\tilde{F}(0)=\left[\begin{array}{ccc}
0 & 0 & ke^{i\theta}\\
0 & 0 & 0\\
ke^{-i\theta} & 0 & 0
\end{array}\right]
\end{equation}
where the factor of $\sqrt{2}$ is to ensure that the normalisation
is correct. Let us consider further the boundary conditions on the
postulated Hamiltonian. As the Hamiltonian is manifestly periodic,
we must have $\tilde{H}(t+T)=\tilde{H}(t)$. If we expand the cosine
and sine arguments in the matrix above using the sum of angles formulae,
one immediately obtains that $\sin kT=0$ and $\cos kT=1$. We shall
demonstrate that determining the nature of the relationship between
the variables $R$ and $k$ is key to understanding this dynamical
system.

\section{Time Evolution Operators I}

Consider for now an arbitrary Hamiltonian similar to what we have
been working with previously. One may write down the following transformation
law which takes the Hamiltonian to the diagonal representation $\tilde{H}=\hat{Q}\hat{L}\hat{Q}^{\dagger}$.
Writing it all out explicitly, we have for the matrices composing
the decomposition:

\begin{equation}
\tilde{H}=\left[\begin{array}{ccc}
0 & \varepsilon_{1} & 0\\
\overline{\varepsilon}_{1} & 0 & \varepsilon_{2}\\
0 & \overline{\varepsilon}_{2} & 0
\end{array}\right]
\end{equation}
\begin{equation}
\hat{L}=\left[\begin{array}{ccc}
R & 0 & 0\\
0 & -R & 0\\
0 & 0 & 0
\end{array}\right]
\end{equation}
\begin{equation}
\hat{Q}=\left[\begin{array}{ccc}
\dfrac{\varepsilon_{1}}{\sqrt{2}R} & -\dfrac{\varepsilon_{1}}{\sqrt{2}R} & -\dfrac{\varepsilon_{2}}{R}\\
\dfrac{1}{\sqrt{2}} & \dfrac{1}{\sqrt{2}} & 0\\
\dfrac{\overline{\varepsilon}_{2}}{\sqrt{2}R} & -\dfrac{\overline{\varepsilon}_{2}}{\sqrt{2}R} & \dfrac{\overline{\varepsilon}_{2}}{R}
\end{array}\right]
\end{equation}
where we implicitly assume that $\left|\varepsilon_{1}(t)\right|^{2}+\left|\varepsilon_{2}(t)\right|^{2}=R^{2}$.
We shall not assume that this decomposition is unique, and indeed
this will not turn out to be the case. We may formulate an operator
that describes the change from state to state in time via the composition
law $\hat{U}(t,s)=\hat{Q}(t)\hat{Q}^{\dagger}(s)$ uniqueness nothwithstanding.
This decomposition is however unitary, so we may use the dagger symbol
interchangeably with the inverse sign. Using this, we obtain the following
formula: 
\begin{equation}
\hat{U}(t,s)=\dfrac{1}{R^{2}}\left[\begin{array}{ccc}
\varepsilon_{+}(t,s) & 0 & \varepsilon_{-}(t,s)\\
0 & R^{2} & 0\\
-\bar{\varepsilon}_{-}(t,s) & 0 & \bar{\varepsilon}_{+}(t,s)
\end{array}\right]
\end{equation}
where $\varepsilon_{+}(t,s)=\varepsilon_{1}(t)\bar{\varepsilon}_{1}(s)+\varepsilon_{2}(t)\bar{\varepsilon}_{2}(s)$
and $\varepsilon_{-}(t,s)=\varepsilon_{1}(t)\varepsilon_{2}(s)-\varepsilon_{2}(t)\varepsilon_{1}(s)$.
We already know the form of the control fields from the previous calculation,
so by substitution we obtain the following formula: x
\begin{equation}
\hat{U}(t,s)=\left[\begin{array}{ccc}
\cos(k(t-s)) & 0 & ie^{-i\theta}\sin(k(t-s))\\
0 & 1 & 0\\
ie^{i\theta}\sin(k(t-s)) & 0 & \cos(k(t-s))
\end{array}\right]
\end{equation}
which conforms to the time-translation invariance principle $\hat{U}(t,s)=\hat{U}(t-s,0)$
and is unitary. Now, this is not the only unitary transformation one
may find on this space, but it is a simple exercise to show that it
achieves the purpose of enabling the diagonalisation of the time evolution
operator as follows. We can write the decomposition as thus: 
\begin{equation}
\tilde{H}(t)=\hat{Q}(t)\hat{L}\hat{Q}^{\dagger}(t)
\end{equation}
where by construction we have $i\hat{\dot{Q}}=\tilde{H}\hat{Q}=\hat{L}\hat{Q}$,
and also$-i\hat{\dot{Q}}^{\dagger}=\hat{Q}^{\dagger}\tilde{H}=\hat{Q}^{\dagger}\hat{L}$.
We may then write
\begin{equation}
\tilde{H}(t)=\hat{Q}(t)\hat{Q}^{\dagger}(s)\hat{Q}(s)\hat{L}\hat{Q}^{\dagger}(s)\hat{Q}(s)\hat{Q}^{\dagger}(t)
\end{equation}
\begin{equation}
=\hat{U}(t,s)\tilde{H}(s)\hat{U}^{\dagger}(t,s)
\end{equation}
which we may explicitly compute to check that, for our chosen initial
condition on the Hamiltonian, it does indeed produce the required
output.

\section{Conservation Law}

It is obvious by now that the constraint, and its constancy, is primary
in understanding this system. What is required is an understanding
of how the dynamics is affected by the constraint. We know that it
is a constant, but we do not know that it is zero. It must therefore
have measurable effects and be present directly in the equations of
motion. We will now show how to achieve this in a most salient fashion.
We know, by the nature of the problem we are considering, that the
constraint is invariant to the time evolution operator via $\hat{U}(t,0)\tilde{F}_{0}\hat{U}^{\dagger}(t,0)=\tilde{F}_{0}$.
We can rewrite this as $\tilde{F}_{0}\hat{U}-\hat{U}\tilde{F}_{0}=\left[\tilde{F}_{0},\hat{U}\right]=0$.
Let us see what effect this has on the dynamics of state. Writing
the equation of motion for the time evolution operator, we find:
\begin{equation}
i\dfrac{d\hat{U}}{dt}=\tilde{H}(s)\hat{U}+\left[\tilde{F}_{0},\hat{U}\right]
\end{equation}
which we are allowed to do as it is just adding zero to the right
hand side of the equation. Writing this out in full, we now have the
expression:
\begin{equation}
i\dfrac{d\hat{U}}{dt}=(\tilde{H}(s)+\tilde{F}_{0})\hat{U}-\hat{U}\tilde{F}_{0}
\end{equation}
What is happening here is the combination of two conservation laws
which must be obeyed. The existence of the Schrödinger equation for
the time evolution operator implies the existence of discrete energy
eigenstates, which is equivalent to a rule for quadratures via the
isotropic constraint on the Hamiltonian. The constraint in this physical
system also has a quadrature sum rule, implied by the invariance under
the time evolution operator. The physical system then must be configured
with both conservation laws adding up vectorially, or in this case
via a phasor sum rule over complex matrices. With this in mind, we
can now immediately write down the time evolution operator for the
quantum state including the constraint:
\begin{equation}
\hat{U}(t,0)=\hat{U}_{1}\hat{U}_{2}
\end{equation}
\begin{equation}
\hat{U}_{1}=\exp\left(-i\left[\tilde{F}_{0}t+\int_{0}^{t}\tilde{H}(s)ds\right]\right)
\end{equation}
\begin{equation}
\hat{U}_{2}=\exp\left(i\tilde{F}_{0}t\right)
\end{equation}
This is the forward-in-time equation. We now see how each piece fits
together, and how everything works in this complex system. We now
have the unitary operator which most generally describes the motion
of the system. This will naturally imply certain symmetries depending
on the output of $\left[\tilde{H}(s),\tilde{F}_{0}\right]$. In this
case, we know that it is not zero. Simple computation gives:
\[
\left[\tilde{H}(s),\tilde{F}_{0}\right]=iR\left[\begin{array}{ccc}
0 & -\sin ks & 0\\
-\sin ks & 0 & -ie^{-i\theta}\cos ks\\
0 & ie^{i\theta}\cos ks & 0
\end{array}\right]
\]
\begin{equation}
=i\tilde{H}(s-\dfrac{\pi}{2k})
\end{equation}
The formula above demostrates explicitly the back-action caused by
the interaction between the Hamiltonian and constraint. If we were
to expand the first exponential in the time evolution operator, one
is free to move the constraint inside the integral sign as it is a
constant matrix. One would then be faced with a series of integrals
over intermediate times in order to evaluate the system. We shall
not take this approach, that is another problem in and of itself.
The correct way to proceed here is to look at the behaviour of each
of the constituent unitary operators in the time evolution operator
over a whole cycle. We then can simply extrapolate backwards using
the composition law of the time evolution operator to get the value
at any other time.

\section{Floquet Representation}

Reconsidering the situation, we now are presented with calculating
the operator $\hat{U}_{1}$ from above. It seems to be worse than
before. However, it may be resolved by resorting to the periodic nature
of the known Hamiltonian. We know from the physics that it must also
obey a time composition law; this is implied from the periodicity
of the system. So if we know the value of the operator at the conclusion
of a cycle, we can work backwards to find the value at any other time
between the intial time and the terminal value via
\begin{equation}
\hat{U}_{1}(T+t,T+s)=\hat{U}_{1}(t,s)
\end{equation}
\begin{equation}
\hat{U}_{1}(T-t,0)=\left\{ \begin{array}{c}
\hat{U}_{1}(T,t)\\
\hat{U}_{1}(T,0)\hat{U}_{1}^{\dagger}(t,0)\\
\hat{U}_{1}(T,0)\hat{U}_{1}(-t,0)
\end{array}\right.
\end{equation}
Calculating the exponential operator for a whole period, we can write:
\begin{equation}
\hat{U}_{1}(T,0)=\exp\left(-i\left[\int_{0}^{T}\left(\tilde{H}(s)+\tilde{F}_{0}\right)ds\right]\right)
\end{equation}
which we will write in the form 
\begin{equation}
\hat{U}_{1}(T,0)=\exp\left(-i\hat{B}T\right)
\end{equation}
\begin{equation}
\hat{B}=\dfrac{1}{T}\int_{0}^{T}\left(\tilde{H}(s)+\tilde{F}_{0}\right)ds
\end{equation}
We are now in a position to apply the Floquet theorem directly. We
know that the matrix $\hat{B}$ must be periodic, and there must be
a continuous isometry that takes the unitary operator to a diagonal
representation. Calculating the eigenvalues of the matrix $\tilde{H}(0)+\tilde{F}_{0}$,
we find that we must have a diagonal representation for this unitary
operator of the form: 
\begin{equation}
\hat{U}_{F}(T,0)=\left[\begin{array}{ccc}
e^{-iT\Delta} & 0 & 0\\
0 & e^{iT\Delta} & 0\\
0 & 0 & 1
\end{array}\right]
\end{equation}
where $\Delta=\sqrt{R^{2}+k^{2}}$. In this representation, we must
therefore have: 
\begin{equation}
\hat{U}_{1}(T,0)=\hat{Y}^{\dagger}\hat{U}_{1F}(T,0)\hat{Y}
\end{equation}
with associated isometry operator:
\begin{equation}
\hat{Y}=\left[\begin{array}{ccc}
\dfrac{e^{-i\theta}}{\sqrt{2}} & -\dfrac{e^{-i\theta}}{\sqrt{2}} & 0\\
0 & 0 & 1\\
\dfrac{1}{\sqrt{2}} & \dfrac{1}{\sqrt{2}} & 0
\end{array}\right]
\end{equation}
Writing $\hat{B}$ in matrix form, we have to compute the following:
\begin{equation}
\exp\left(-i\hat{B}T\right)=\hat{Y}^{\dagger}\exp\left(-iT(\hat{Y}\hat{B}\hat{Y}^{\dagger})\right)\hat{Y}
\end{equation}
\begin{equation}
=\hat{Y}^{\dagger}\exp\left(-iT\left[\hat{Y}\tilde{F}_{0}\hat{Y}^{\dagger}+\left(\dfrac{1}{T}\int_{0}^{T}\hat{Y}\tilde{H}(s)\hat{Y}^{\dagger}ds\right)\right]\right)\hat{Y}
\end{equation}
Let us now examine the effect of moving this isometry inside the matrix
exponential. The first part is just some constant matrix. The second
part is effectively another Hamiltonian, in different reference frame.
We have from the previous section an expression for the time evolution
operator. We can write the transformations in the diagonal representation
in the format given below: 
\begin{equation}
\tilde{H}(s)=\hat{U}(s,0)\tilde{H}(0)\hat{U}^{\dagger}(s,0)
\end{equation}
\begin{equation}
\tilde{H}_{F}(s)=\hat{U}_{F}(s,0)\tilde{H}_{F}(0)\hat{U}_{F}^{\dagger}(s,0)
\end{equation}
\begin{equation}
\hat{U}_{1F}(s,0)=\hat{Y}\hat{U}_{1}(s,0)\hat{Y}^{\dagger}
\end{equation}
Our integral then is considerably simplified; we only need to evaluate
the following:
\begin{equation}
\hat{Y}^{\dagger}\exp\left(-iT\left[\hat{Y}\tilde{F}_{0}\hat{Y}^{\dagger}+\left(\dfrac{1}{T}\int_{0}^{T}\tilde{H}_{F}(s)ds\right)\right]\right)\hat{Y}
\end{equation}
and in particular, the expression:
\begin{equation}
\exp\left(-iT\left[\tilde{F}_{0}+\hat{Y}^{\dagger}\left(\dfrac{1}{T}\int_{0}^{T}\tilde{H}_{F}(s)ds\right)\hat{Y}\right]\right)
\end{equation}
We shall abbreviate this operator to be equal to $\exp\left(-iT\left[\tilde{F}_{0}+\hat{Y}^{\dagger}\hat{S}\hat{Y}\right]\right)$where
we have:
\begin{equation}
\hat{H}_{F}(s)=\hat{U}_{F}(s,0)\hat{Q}(0)\hat{L}\hat{Q}^{\dagger}(0)\hat{U}_{F}^{\dagger}(s,0)
\end{equation}
\begin{equation}
\hat{S}=\dfrac{1}{T}\int_{0}^{T}\tilde{H}_{F}(s)ds
\end{equation}
and we can use the unitary operator in the diagonal representation
as displayed before, and the solution matrices as found in the earlier
sections. Writing out the matrices explicitly, and remembering to
transform our initial condition for the Hamiltonian into the Floquet
reference frame, we find: 
\begin{equation}
\hat{H}_{F}(0)=\dfrac{R}{2}\left[\begin{array}{ccc}
0 & 0 & e^{i\theta}\\
0 & 0 & -e^{i\theta}\\
e^{-i\theta} & -e^{-i\theta} & 0
\end{array}\right]
\end{equation}
Hence one may directly evaluate the integral of the matrix using the
equation for the time evolution operator: 
\begin{equation}
\hat{S}=\hat{S}(T)-\hat{S}(0)
\end{equation}
\begin{equation}
S(t)=\dfrac{R}{2\Delta}\left[\begin{array}{ccc}
0 & 0 & e^{i(t\Delta+\theta)}\\
0 & 0 & -e^{-i(t\Delta-\theta)}\\
e^{-i(t\Delta+\theta)} & -e^{+i(t\Delta-\theta)} & 0
\end{array}\right]
\end{equation}
which can be rewritten in the format below:
\begin{equation}
\hat{S}=\dfrac{R}{2\Delta}\left[\begin{array}{ccc}
0 & 0 & z^{\star}\\
0 & 0 & w^{\star}\\
z & w & 0
\end{array}\right]
\end{equation}
We may write a formula for the exponential of a matrix of this format,
where we assume that $\left|z\right|^{2}+\left|w\right|^{2}=1$. For
now, we drop the explicit time dependence on $z$ and $w$ and compute
it naïvely. The method is as follows. Assume for now that $T$ is
just a constant factor. We have the exponent which we can write as:
\begin{equation}
\exp\left(-iT\left[\tilde{F}_{0}+\hat{Y}^{\dagger}\hat{S}\hat{Y}\right]\right)=\exp\left(-i\hat{B}T\right)
\end{equation}
\begin{equation}
\hat{B}=\tilde{F}_{0}+\dfrac{1}{T}\hat{Y}^{\dagger}\hat{S}\hat{Y}
\end{equation}
which we can rewrite as the matrix:
\begin{equation}
\hat{B}=\left[\begin{array}{ccc}
0 & 0 & ke^{i\theta}+\dfrac{R}{\sqrt{2}T\Delta}\overline{z}\\
0 & 0 & \dfrac{R}{\sqrt{2}T\Delta}\overline{w}\\
ke^{-i\theta}+\dfrac{R}{\sqrt{2}T\Delta}z & \dfrac{R}{\sqrt{2}T\Delta}w & 0
\end{array}\right]
\end{equation}

\begin{equation}
=\Delta\left[\begin{array}{ccc}
0 & 0 & \overline{Z}/\Delta\\
0 & 0 & \overline{W}/\Delta\\
Z/\Delta & W/\Delta & 0
\end{array}\right]
\end{equation}
\begin{equation}
\left|Z\right|^{2}+\left|W\right|^{2}=\left(k^{2}+\dfrac{R^{2}}{\Delta^{2}T^{2}}\right)
\end{equation}
Now, we can always renormalise the complex functions in this matrix
by changing the scale of the time so that they obey $\left|Z\right|^{2}+\left|W\right|^{2}=\Delta^{2}$
using the identity above, by choosing values of $R$, $k$ and consequently
$\Delta$ for a given $T$. where $\Phi=T\Delta$. Now, at the moment
$Z$ and $W$ are effectively constants in the complex plane with
an appropriate normalisation. That is the proviso we have been working
with so far. We can renormalise once more via a unitary transformation
and use the following parameterisation: 
\begin{equation}
Z=ke^{-i\theta}
\end{equation}
\begin{equation}
W=R
\end{equation}
and we will be able to maintain $\left|Z\right|^{2}+\left|W\right|^{2}=\Delta^{2}$
for any particular time. The exponential of this matrix is then given
by the formula:
\begin{equation}
\begin{array}{c}
\hat{U}_{1F}(T,0)=\exp\left(-i\hat{B}T\Delta\right)=\\
\left[\begin{array}{ccc}
u & Rk\left(u-1\right)\dfrac{e^{i\theta}}{\Delta^{2}} & -ik\dfrac{e^{i\theta}v}{\Delta}\\
Rk\left(u-1\right)\dfrac{e^{-i\theta}}{\Delta^{2}} & 1 & -iR\dfrac{v}{\Delta}\\
-ik\dfrac{e^{-i\theta}v}{\Delta} & -iR\dfrac{v}{\Delta} & u
\end{array}\right]
\end{array}
\end{equation}
where for convenience we write $\cos T\Delta=u$,$\sin T\Delta=v$.
For now, we will keep this formula for the unitary in the Floquet
picture. To end this task, all we are required to do is calculate
the other unitary $\hat{U}_{2}(t,0)$. Now this matrix is of the form:
\begin{equation}
\hat{U}_{2}=\exp\left(i\tilde{F}_{0}t\right)
\end{equation}
We have already evaluated the more difficult problem of calculating
this matrix with the previous task. However, it is important to remember
that the result we have obtained is the solution that is travelling
forwards in time, and also in the Floquet picture. Inverting the sign
of time in the matrix above and setting all the Hamiltonian coefficients
to zero, we obtain: 
\begin{equation}
\hat{U}_{2F}(T,0)=\left[\begin{array}{ccc}
\cos kT & 0 & ie^{i\theta}\sin kT\\
0 & 1 & 0\\
ie^{-i\theta}\sin kT & 0 & \cos kT
\end{array}\right]
\end{equation}
remembering that we are still in the reference frame where the unitary
operator is diagonal. We recognise this matrix from the previous calculation
involving the matrix of time dependent states. Everything is as it
should be. We therefore can write the time evolution operator in the
form: 
\begin{equation}
\hat{U}(T,0)=\hat{U}_{1}\hat{U}_{2}=\hat{Y}^{\dagger}\hat{U}_{1F}\hat{U}_{2F}\hat{Y}
\end{equation}
Now, we can observe several things. Firstly, although we have treated
$T$, the period time, as a constant, in reality, the implicit parametric
freedom within the system allows us to now see that the calculations
we have carried out apply for any time whatsoever. This is due to
the additive nature of the unitary operator. It doesn't matter where
we start in the cycle, only how far through it we are relative to
the initial point. We also notice the following; that we must have
$\sin T_{0}\Delta=0$ for the actual period time. This gives us the
relationship: 
\begin{equation}
T_{0}=\dfrac{2m\pi}{\Delta}
\end{equation}
from the unitary operator $\hat{U}_{1F}$ and 
\begin{equation}
T_{0}=\dfrac{2n\pi}{k}
\end{equation}
from the operator $\hat{U}_{2F}$ for some integers. In particular,
we have immediately that
\begin{equation}
\dfrac{\Delta}{k}=\dfrac{m}{n}
\end{equation}
We are now in a position to truly understand the dynamics of the time
evolution operator in this space. Since we can derive the composition
formula $\hat{U}_{1}^{n}(\delta t,0)=\hat{U}_{1}(n\delta t,0)=\hat{U}_{1}(T,0)$,
we can see how the dynamics in this representation is developed. We
have, on the one hand, a rotation in space- with phase gathering on
counter-rotating particles. On the other, we have the rotation as
composed above, which we have shown is related to a conservation law
between the Hamiltonian and constraint. We can see that the periodicity
of this operator changes as a function of the ratio between the control
and Hamiltonian field strengths, and that for periodicity to occur,
we must have that $\dfrac{\Delta}{k}=\dfrac{m}{n}$ . As the space
is not simply connected, we have to make sure that we preserve a certain
group of fundamental rotations as the particle travels through on
its periodic trajectory. The constraint gives up energy to the system,
the system absorbs it, then transmits it back to the constraint. At
all points we must obey the correct quadrature rules and make sure
that the summation is correct to ensure that the conservation laws
of energy and angular momentum are obeyed. It is curious that such
a simple premise can lead to such large conclusions. To finish this
calculation, let us determine the normalisation of the $Z$ and $W$.
We have, as before, the conservation law:
\begin{equation}
\left|Z\right|^{2}+\left|W\right|^{2}=\left(k^{2}+\dfrac{R^{2}}{\Delta^{2}T^{2}}\right)
\end{equation}
If we square the relationship relating the periodicities of the two
unitary operators, we find:
\begin{equation}
\dfrac{\Delta^{2}}{k^{2}}=\dfrac{k^{2}+R^{2}}{k^{2}}=\dfrac{m^{2}}{n^{2}}
\end{equation}
Therefore, after re-arrangement, we have the identity:
\begin{equation}
R^{2}=\left(\dfrac{m^{2}}{n^{2}}-1\right)k^{2}
\end{equation}
Now the left hand side of this identity is a positive number. We must
therefore have $m=2$, $n=1$. Completing, we find directly that $R=\sqrt{3}k$.
The normalisation of the conservation law is then given by: 
\begin{equation}
k^{2}+\dfrac{R^{2}}{\Delta^{2}T^{2}}=k^{2}+R^{'2}
\end{equation}
There is a factor of $2\pi$ that is cancelled out here. The theory
calculated has the property that, as long as we renormalise $R$ and
thereby $\Delta$, the physics is unchanged. In this case, we are
looking at the new transformed reference frame where $R'=2\pi R$.
The normalisation will then be given correctly as above. This is related
to the spherical nature of the isotropic condition. Therefore the
right hand side of the equation above can be written as:
\begin{equation}
\Delta=2k
\end{equation}
We have covered the whole geometry of this space in tackling this
task. It has been a monumental effort. In doing so, we have addressed
many outstanding questions about these groups.

\section{Periodicity on SU(3)}

We shall now examine the periodic behaviour of the Hamiltonian and
state on this complex space. It has already proven much more difficult
to treat than examples on SU(2), where we have access to a familiar
geometric arrangement of spherical geometry. It is even more difficult
to handle than SU(4), in that we do not have access to any block-diagonal
matrix methods. Here we are confronted with the true complexity of
quantum states. As the state meanders in its periodic trajectory,
it is reasonable to look at what other features are visited at extremums
of the periodic functions. One can write down a diagram to represent
this:
\begin{equation}
\begin{array}{ccc}
\left[\begin{array}{ccc}
0 & 1 & 0\\
1 & 0 & 0\\
0 & 0 & 0
\end{array}\right] & \rightarrow & \left[\begin{array}{ccc}
0 & 0 & 0\\
0 & 0 & -ie^{-i\theta}\\
0 & ie^{i\theta} & 0
\end{array}\right]\\
\uparrow & \tilde{H}(t) & \downarrow\\
\left[\begin{array}{ccc}
0 & 0 & 0\\
0 & 0 & ie^{-i\theta}\\
0 & -ie^{i\theta} & 0
\end{array}\right] & \leftarrow & \left[\begin{array}{ccc}
0 & -1 & 0\\
-1 & 0 & 0\\
0 & 0 & 0
\end{array}\right]
\end{array}
\end{equation}
Each of these motions is happening over an equal interval during a
period $T$, so we immediately know that we must have the expression
$kT=2\pi$, and $\tilde{H}(t+T)=\tilde{H}(t)$. Let us now consider
the periodicity of the state. Using the unitary operator developed
in the previous section, we can write the state at time $t$ as $\left|\psi(t)\right\rangle =\hat{U}(t,0)\left|\psi(0)\right\rangle $.
Let's assume for now that we have an unspecified initial condition.
We obtain:
\begin{equation}
\left|\psi(t)\right\rangle =\left[\begin{array}{ccc}
\cos(kt) & 0 & ie^{-i\theta}\sin(kt)\\
0 & 1 & 0\\
ie^{i\theta}\sin(kt) & 0 & \cos(kt)
\end{array}\right]\left[\begin{array}{c}
C_{1}\\
C_{2}\\
C_{3}
\end{array}\right]
\end{equation}
\begin{equation}
=\left[\begin{array}{c}
C_{1}\cos(kt)+ie^{-i\theta}C_{3}\sin(kt)\\
C_{2}\\
ie^{i\theta}C_{1}\sin(kt)+C_{3}\cos(kt)
\end{array}\right]
\end{equation}
By observation, we also have similar periodicity $\left|\psi(2\pi n/k)\right\rangle =\left|\psi(nT)\right\rangle =\left|\psi(0)\right\rangle $.
Finally, we may examine also the unitary operator $\hat{U}(t,s)$.
\begin{equation}
\hat{U}(t,s)=\left[\begin{array}{ccc}
e^{i\Delta(t-s)} & 0 & 0\\
0 & e^{-i\Delta(t-s)} & 0\\
0 & 0 & 1
\end{array}\right]
\end{equation}
By substitution, one immediately gains the periodicity law for this
operator $\hat{U}(t+T,s+T)=\hat{U}(t,s)$. We know from the calculation
in the previous section that we must therefore have $\Delta$ to be
in proportion to that of $k$. The exact nature of this relationship
will be examined in the next section. For now, we just establish this
side of what is hoped to be a double-sided equality.

\section{Time Evolution Operators II}

There is another approach we can use to solve the Schrödinger equation
in this situation. To do this, we shall look at the time evolution
of the Hamiltonian operator. We know, for a fact, that the constraint
matrix is a constant. It therefore must not evolve in time. We have
the time evolution equation:

\begin{equation}
\hat{U}(t,0)\left[\tilde{H}_{0}+\tilde{F}_{0}\right]\hat{U}^{\dagger}(t,0)=\tilde{H}(t)+\tilde{F}_{0}
\end{equation}
where we use the subscript to accentuate the fact that these operators
are constant matrices and not matrix variables. Using $\hat{U}\hat{U}^{\dagger}=\mathbf{1}$,
and $i\dfrac{d\hat{U}}{dt}=\tilde{H}(t)\hat{U}$, we may derive the
equation for the time evolution operator in the following form, under
the given proviso of a constant constraint: 
\begin{equation}
i\dfrac{d\hat{U}}{dt}=\hat{U}\left[\tilde{H}_{0}+\tilde{F}_{0}\right]-\tilde{F}_{0}\hat{U}
\end{equation}
We can therefore write the solution for the time evolution operator
by the following product of exponentials:
\begin{equation}
\hat{U}(t,0)=\exp\left(i\tilde{F}_{0}t\right)\exp\left(-i\left[\tilde{H}_{0}+\tilde{F}_{0}\right]t\right)=\hat{U}_{+}\hat{U}_{-}
\end{equation}
This technique will only ever work in the scenario where we have a
constant constraint. Evaluating the matrix exponentials, we obtain
the following:
\begin{equation}
\hat{U}_{-}(t,0)=\exp\left(-it\hat{A}\right)
\end{equation}
\begin{equation}
=\left[\begin{array}{ccc}
\cos\Phi & -i\dfrac{R\sin\Phi}{\Delta} & -\dfrac{ike^{-i\theta}\sin\Phi}{\Delta}\\
-i\dfrac{R\sin\Phi}{\Delta} & \dfrac{k^{2}+R^{2}\cos\Phi}{\Delta^{2}} & \dfrac{kRe^{-i\theta}(\cos\Phi-1)}{\Delta^{2}}\\
-\dfrac{ike^{i\theta}\sin\Phi}{\Delta} & \dfrac{kRe^{i\theta}(\cos\Phi-1)}{\Delta^{2}} & \dfrac{R^{2}+k^{2}\cos\Phi}{\Delta^{2}}
\end{array}\right]
\end{equation}
\begin{equation}
\hat{A}=\left[\begin{array}{ccc}
0 & R & ke^{-i\theta}\\
R & 0 & 0\\
ke^{i\theta} & 0 & 0
\end{array}\right]
\end{equation}
and $\Phi=t\Delta$, $\Delta=\sqrt{k^{2}+R^{2}}$. Note the similarity
to the unitary operator in the previous section. The other unitary
is easily evaluated from the formula above; one just takes the coefficient
$R$ as zero, and inverts the time via $t\rightarrow-t$. The matrix
obtained is thus:
\begin{equation}
\hat{U}_{+}(t,0)=\left[\begin{array}{ccc}
\cos kt & 0 & ie^{-i\theta}\sin kt\\
0 & 1 & 0\\
ie^{i\theta}\sin kt & 0 & \cos kt
\end{array}\right]
\end{equation}
which we also recognise from the previous calculation. We are nearly
done. All the pieces of the puzzle in this hypercomplex space are
together, the pieces slightly less unfamiliar than when we arrived.
We may now establish the boundary conditions using the unitary operator.
We must have that $\hat{U}(T,0)=\mathbf{1}$. Observing the two unitary
operators that compose the total time evolution operator, we find
immediately that we must have both $\sin kT=0$, implying that $kT=2n\pi$
as before, and also that $\sin T\Delta=0$, giving $T\Delta=2m\pi$
additionally. Moreover, this immediately gives us a physical understanding
as to why the rotation was so complicated in the previous section.
Once again, we follow the same argument, observing that we must have,
by observation, the relationship $\dfrac{\Delta}{k}=\dfrac{m}{n}$,
where $m,n$ are integers greater than zero. Following the argument
as in the preceding sections, one can see that the lowest possible
solution that can be obtained is $m=2,n=1$. Inserting this into the
expression, we find that $R=\sqrt{3}k$, and hence 
\begin{equation}
\dfrac{\Delta}{k}=\dfrac{\sqrt{k^{2}+3k^{2}}}{k}=2
\end{equation}
which satisfies our need for multiple periodicity as per requirements.
This result is slightly different to that obtained in \cite{Morrison 1,Morrison 3}
as we are considering total periodicity to return to an initial state,
not just state-to-state transfer. It seems very strange and unphysical
to be multiplying by such numerical factors. However, it determines
the outcome. One may find a sequence of such numbers in varying ratios,
that seem to be related to a declination in some plane within the
space relative to some axial element. One way to view the situation
is of complex functions in stacked planes that carve out conic sections
as they transcribe their orbits in opposite directions, however a
clear geometric sketch remains elusive.

\section{Constraint Reconsidered}

Now that we have a fair degree of confidence in handling this complex
dynamic system, let us look at some of our earlier hypotheses and
see how they develop in light of our increased understanding of this
strange and complicated geometric space. Consider the constraint $\hat{F}_{0}$.
Initially, we assumed that it was acceptable to take one of the generators
in the matrix to be equal to zero, as it was commutative with every
other member of the group. We also calculated that the other diagonal
element in the postulated constraint was zero; the reasons for this
have become apparent. As this element of the group has played such
a central role in the calculation preceding, and obviously defines
all aspects of the group, while being part of the minimal commutative
subgroup, it is obviously the centralising element of the algebra.
Let us break this assumption, and see what results were they not zero,
but the Hamiltonian and unitary operators are as above. We know that,
regardless, the constraint will evolve in time via $\hat{U}(t,0)\tilde{F}(0)\hat{U}^{\dagger}(t,0)=\tilde{F}(t)$.
Substitution of the matrix:

\begin{equation}
\tilde{F}_{0}=\left[\begin{array}{ccc}
\omega_{1} & 0 & ke^{i\theta}\\
0 & -(\omega_{1}+\omega_{2}) & 0\\
ke^{-i\theta} & 0 & \omega_{2}
\end{array}\right]
\end{equation}
into the time evolution equation, and using the operator for forward-in-time
translation, we get: 
\begin{equation}
\tilde{F}(t)=\left[\begin{array}{ccc}
\varLambda_{1}\cos^{2}kt-\varLambda_{2} & 0 & e^{i\theta}\Xi\\
0 & \omega_{2} & 0\\
-e^{-i\theta}\Xi^{\star} & 0 & -\varLambda_{1}\cos^{2}kt+\varLambda_{2}
\end{array}\right]
\end{equation}
\begin{equation}
2\omega_{1}+\omega_{2}=\varLambda_{1}
\end{equation}
\begin{equation}
\omega_{1}+\omega_{2}=\varLambda_{2}
\end{equation}
\begin{equation}
\Xi=\dfrac{1}{2}\left(k\cos2kt+i\omega_{2}\sin2kt\right)
\end{equation}
A simple observation shows that, conditional on our having calculated
the time optimal Hamiltonian, that the diagonal elements must be zero
in the above formula. However, if they are not, what happens is that
transitions on the off-diagonal elements get over-cycled, and the
path becomes no longer geodesic. We shall now show that we have actually
solved another problem in tackling the more difficult task of finding
the time evolution operators and considering the full time dependence
of a complicated Hamiltonian. We can invert the task, and look at
how the problem would change if one was to take the Hamiltonian as
constraint and vice versa. The Hamiltonian can then be rewritten in
the form $\tilde{H}_{P}=\hat{P}\tilde{H}\hat{P}^{\dagger}=\mu(\sigma\cdot\mathbf{B})$.
\begin{equation}
\tilde{H}=\left[\begin{array}{ccc}
\omega_{1} & 0 & \kappa\\
0 & 0 & 0\\
\overline{\kappa} & 0 & -\omega_{1}
\end{array}\right]
\end{equation}
 Writing the constraint out in matrix form, and maintaining the structure
of the original algebra, we have:
\begin{equation}
\tilde{F}_{P}=\hat{P}\tilde{F}\hat{P}^{\dagger}=\left[\begin{array}{ccc}
\omega_{2} & \varepsilon_{1} & 0\\
\overline{\varepsilon}_{1} & -2\omega_{2} & \varepsilon_{2}\\
0 & \overline{\varepsilon}_{2} & \omega_{2}
\end{array}\right]
\end{equation}
as the commuting element is invariant to the permutation transform.
The parameters in the constraint might change label, however this
shall prove irrelevant. We know from the quantum brachistochrone equation
that this $\tilde{H}$ will be a constant of the motion under the
chosen constraint. We therefore can immediately evaluate the time
evolution operator via the Cayley-Hamilton theorem, viz. 
\begin{equation}
\exp\left(-it\tilde{H}\right)=\mathbf{1}-i\tilde{H}\dfrac{\sin\nu t}{\nu}+(\cos\nu t-1)\dfrac{\tilde{H}^{2}}{\nu^{2}}
\end{equation}
where $\nu^{2}=\omega_{1}^{2}+\left|\kappa\right|^{2}$ from which
we can find easily evolve a state to find e.g.
\begin{equation}
\left|\psi\right\rangle =\left[\begin{array}{c}
\cos\nu t-i\dfrac{n_{z}\sin\nu t}{\nu}\\
0\\
-i\dfrac{\overline{\kappa}}{\nu}\sin\nu t
\end{array}\right]
\end{equation}
What we are seeing here is another reflection of a conservation law
emerging from the dynamics. In this case, the fact that $\omega_{1}^{2}+\left|\kappa\right|^{2}$
equals a constant. We have implicitly assumed this through the dynamical
system that we are operating under with the isotropic constraint.
This quantity represents the quantity that is conserved through the
motion, which is related to the angular momentum of the system. The
complexities that have arisen are due to the consideration of the
system and the constraint at the same time, and ensuring that all
conserved quantities preserve the periodicity as well. In this case,
it just happens to be a particularly simple system to solve for the
Hamiltonian dynamics, and the constraint is complicated vis a vis
our earlier difficulties.

\section{Eigenspace Degeneracy}

We must now comment on how the essential degeneracy in the eigenspace,
which has caused so much difficulty, can be shown most clearly. What
we shall do is demonstrate a series of matrix transformations, all
of which map to the same degenerate, non-invertible operator in a
unitary way. To demonstrate this, we list the following isometric
transformations, which may be easily derived using the matrices in
Appendix A:
\begin{equation}
\hat{X}_{Q}\hat{L}\hat{X}_{Q}^{\dagger}=\left[\begin{array}{ccc}
0 & \cos t & 0\\
\cos t & 0 & -ie^{-i\theta}\sin t\\
0 & ie^{i\theta}\sin t & 0
\end{array}\right]
\end{equation}
\begin{equation}
\hat{X}_{J}\hat{L}\hat{X}_{J}^{\dagger}=\left[\begin{array}{ccc}
0 & \cos t & 0\\
\cos t & 0 & -i\sin t\\
0 & i\sin t & 0
\end{array}\right]
\end{equation}
\begin{equation}
\hat{X}_{D}\hat{L}\hat{X}_{D}^{\dagger}=\left[\begin{array}{ccc}
0 & -i\cos t & 0\\
i\cos t & 0 & -i\sin t\\
0 & i\sin t & 0
\end{array}\right]
\end{equation}
The first matrix transformation is familiar; it is related to the
transformations we have been studying earlier. The Hamiltonian matrix
that is a product of the decomposition related to $\hat{X}_{D}$ is
related to a question of the Frenet-Serre curve on SU(3). It has properties
somewhat dissimilar in behaviour, so the fact that these two seemingly
disparate systems can be related by a unitary transformation due to
the essential degeneracy of the space is remarkable. This will obviously
reflect itself in ways such as the decay of high-energy particles
where one decay pathway is favoured over another. For now, it is enough
to note that the fact that we may transform unitarily between these
systems by composing the matrices above requires certain symmetries
on the time evolution operator which governs the space. We first consider
the various transformations on the column vectors of each of these
matrix operators. We can use the double angle formulae to expand the
vectors which make up the solution matrices to find linear transforms
on this space: 
\begin{equation}
\hat{X}_{A}=\left[\begin{array}{ccc}
\vdots & \vdots & \vdots\\
a_{1} & a_{2} & a_{3}\\
\vdots & \vdots & \vdots
\end{array}\right]
\end{equation}
 
\begin{equation}
\hat{R}(\sigma)\left|a_{i}(t)\right\rangle =\left|a_{i}(t+\sigma)\right\rangle 
\end{equation}
We obtain the set of unitary operators contained in Appendix II. We
note that the set given by the $\hat{R}_{i,j}(\sigma)$ has the following
property. The matrices that compose this group are very similar to
the unitary operators we have considered in the previous sections.
In particular, $\hat{U}_{+}(t,0)$ in section V, also $\hat{U}(t,s)$
in section II. The picture of what the unitary operators we have calculated
are achieving is more physically accessible. What we have is a rotation
induced by the constraint, and a back-action of the constraint coupled
with the initial Hamiltonian which rotates backwards in time. For
the Hamiltonian to be consistent, and the state periodic, we must
have the physics as described in previous sections. 

We can see from the above, that given we have solved the most generalised
form of this problem, how one may use the unitary transforms to turn
one problem into another by using the essential degeneracy of the
state. We may take the phase in the Hamiltonian we have been working
with as zero to drop down into the first subspace. The second subspace
is the real projection of this dynamical system which can be reached
by another unitary transformation of the complex subspace, which considerably
simplifies the situation. It is possible to see how one could use
this property to turn any set of Hamiltonians which share a set of
eigenvalues into one another. We then only have to solve the problem
in detail for a single Hamiltonian in order to solve for the group
of systems. We shall consider this question later when we discuss
classification of Hamiltonian matrix systems of various types by the
roots of their constituent characteristic polynomials. Let us now
consider another aspect of the degeneracy of this system. We know
that there should be an equivalent of the quantum Fourier transform
on this space. This forms a necessary and integrated part of the dynamics
of the space, we shall see. There are some other related transformations
that can be formed, assume for now we have a complex number $w=e^{i\theta}$
for some arbitrary phase. We can write the following formulae, where
for now we drop constant scale factors.

\begin{equation}
\left\{ \begin{array}{c}
\tilde{H}_{1}=\hat{\Pi}_{1}\hat{L}\hat{\Pi}_{1}^{\dagger}\\
\tilde{H}_{2}=\hat{\Pi}_{2}\hat{L}\hat{\Pi}_{2}^{\dagger}\\
\tilde{H}_{3}=\hat{\Pi}_{3}\hat{L}\hat{\Pi}_{3}^{\dagger}
\end{array}\right\} 
\end{equation}
We give the matrices and associated formulae in the appendix. As the
transformation operators $\hat{\Pi}_{1},\hat{\Pi}_{2},\hat{\Pi}_{3}$
are not unitary, we are at a curious juncture. It looks like this
operator $\hat{L}$ is a non-unitary isometry of another set of matrices.
Let us now expand the arguments of the isometric transformation. We
obtain the following: 
\begin{equation}
\hat{H}_{1}=\tilde{H}_{0}-\tilde{V}(w)
\end{equation}
where $\tilde{V}=\tilde{V}_{sym}+i\tilde{V}_{asym}$. We have written$w=e^{i\theta}$.
However, we may now look at this as the $\tilde{V}(w)$ dependencies
being purely of the phase. We then would obtain $\hat{H}_{1}=\tilde{H}_{0}-\tilde{V}(t)$
where $\theta=\omega t$, by expanding the exponentials. Therefore,
we are now in a situation where we now have: 
\begin{equation}
\tilde{V}(t)=\tilde{H}_{0}-\hat{\Pi}_{1}(t)\hat{L}\hat{\Pi}_{1}^{\dagger}(t)
\end{equation}
From this perspective, what appears to a be time dependent Hamiltonian
$\tilde{V}(t)$ is actually the difference between a static term $\tilde{H}_{0}$
and the time dependent, non-unitary transformation of the centralising
element. This situation is complicated further by the fact that we
can demonstrate different transformations that map to an identitical
subspace. For example, using $\hat{\Pi}_{2}$, $\hat{\Pi}_{3}$from
the appendices we can show that $\hat{\Pi}_{2}\hat{L}\hat{\Pi}_{2}^{\dagger}=\hat{\Pi}_{3}\hat{L}\hat{\Pi}_{3}^{\dagger}$
map to an identical subspace, shown below: 
\begin{equation}
\tilde{H}=\left[\begin{array}{ccc}
0 & 1 & 1\\
1 & 0 & 1\\
1 & 1 & 0
\end{array}\right]-\left[\begin{array}{ccc}
0 & w^{-1} & w^{-2}\\
w & 0 & w^{-1}\\
w^{2} & w & 0
\end{array}\right]
\end{equation}
So this space is essentially complicated by the existence of these
entities on the subspaces, which have singular forms. We must at this
point present the correct form for the qutrit Fourier transform. If
we take cube roots of unity, we may form a group using $z=-\dfrac{1}{2}(1-i\sqrt{3})$,
it is a simple exercise to show that we have $z=\left(z^{\star}\right)^{2}$,
also $z^{2}=z^{\star}$. We may write the discrete Fourier transform
in the following format:
\begin{equation}
\hat{\Pi}=\dfrac{1}{\sqrt{3}}\left[\begin{array}{ccc}
1 & 1 & 1\\
1 & z & z^{2}\\
1 & z^{2} & z^{4}
\end{array}\right]
\end{equation}
By calculation, one can show that this matrix is unitary $\hat{\Pi}\hat{\Pi}^{\dagger}=\hat{\Pi}^{\dagger}\hat{\Pi}=\mathbf{1}$.
However, it is not orthogonal. In fact, one can produce a useful quantum
gate via:
\begin{equation}
\hat{\Pi}^{T}\hat{\Pi}=\hat{\Pi}\hat{\Pi}^{T}=\left[\begin{array}{ccc}
1 & 0 & 0\\
0 & 0 & 1\\
0 & 1 & 0
\end{array}\right]
\end{equation}
We can track this essential difference in the algebra that defines
this sort of structure to one central property. In systems without
degeneracy such as SU(2) we have multiplication rules for spinors
of the form $(\mathbf{a}.\vec{\mathbf{\sigma}})(\mathbf{b}.\vec{\mathbf{\sigma}})=(\mathbf{a}.\mathbf{b})\mathbf{1}+i(\mathbf{a}\times\mathbf{b}).\vec{\mathbf{\sigma}}$.
For SU(4), we have a rule which can be written as $\dfrac{1}{2}\left\{ \mathbf{p},\mathbf{q}\right\} =(\mathbf{p}.\mathbf{q})\mathbf{1}$,
examined in \cite{Morrison 2,Morrison 3}. We may look at this even
more closely and say that the nature of the space is defined by the
way in which a spinor multiplies itself. For SU(2), we have $(\mathbf{n}.\sigma)(\mathbf{n}.\sigma)=\left|\mathbf{n}\right|^{2}\mathbf{1}$
as $\mathbf{n}\times\mathbf{n}=0$; for SU(4), this essential element
is given by the polarisation matrices which obey $(\mathbf{\epsilon}^{\dagger}.\mathbf{p})(\mathbf{\mathbf{\epsilon}}.\mathbf{p})=\left|\mathbf{p}\right|^{2}\mathbf{1}$.
In our system, we have the situation whereby the angular momentum
relations, which play the role of spinors $\mathbf{\epsilon}^{\dagger},\sigma$
in this space, are given by the matrices $\hat{L}_{x}$, $\hat{L}_{y}$,
and $\hat{L}_{z}$ provided in the appendix. Everything about the
algebra is the same, except with the following major and concisive
difference. When we construct an invariant in SU(2) or SU(4), we receive
an identity matrix multiplied by some scalar function. The way in
which this is achieved is through calculation of the squared angular
momentum $\hat{L}^{2}=\sum\hat{L}_{i}^{2}$, which we can evaluate
using the formulae above, with some trickiness in the relativistic
case, but still possible. However, if we try to go about calulating
a similar invariant for this system, we find the following identity:
\begin{equation}
\hat{L}^{2}=3\left[\begin{array}{ccc}
1 & 0 & 0\\
0 & 0 & 0\\
0 & 0 & 1
\end{array}\right]
\end{equation}
Now, this is still an invariant of the system although it is not invertible.
This means that it isn't of any use to find the time evolution operator.
Even worse than this, the case of multiplication of a spinor with
itself now takes the form: 
\begin{equation}
(\mathbf{n}.\mathbf{\hat{L}})(\mathbf{n}.\mathbf{\hat{L}})=\left|\mathbf{n}\right|^{2}\dfrac{\hat{L}^{2}}{3}
\end{equation}
and $\hat{L}^{2}$ is a non-invertible matrix. This is the reason
for all the problems with many valuedness, and the difficulty in calculating
the time evolution operators for this system. In some sense, what
we have done is look at the combined energy from the system and the
constraint as a conjoined object. This has given us enough constants
of motion in order to solve the system, enabling us to bypass the
difficulties we can see above.

The crucial difference in behaviour in these types of quantum systems
will be down to the nature of this angular momentum operator. Because
of this, the naïve implementation of the Floquet theorem actually
fails. The direct application of Floquet theory works perfectly in
SU(2), less perfectly but still functionally in SU(4), this is due
to the simply connected nature of the space. However, as we have shown
in this paper, the spaces we have examined are quite different in
their behaviour. We can see this reflected in progressive degrees
of complication involved in extraction of the unitary operators that
define the movement of time within each of these systems. To invert
the direction of time, we must also incorporate a transformation that
describes the inversion of the direction. In this way, we can most
clearly see the asymmetry of time when it comes to these fundamental
particles. This transformation is a parity inversion and a phase rotation
in opposite directions. By applying this transformation after reversing
the direction of time and changing antiparticles into particles we
will stay invariant.

\section{Classification by Root Systems}

We have touched upon the nature of the polynomial root sets briefly.
Let us now quantify our earlier statements. We can write the characteristic
polynomial for any Hamiltonian that is a $3\times3$ matrix in the
form:
\begin{equation}
\tilde{H}^{3}-\tilde{H}^{2}\textrm{Tr}\left[\tilde{H}\right]-\Delta E^{2}\tilde{H}-\mathbf{1}\det\left[\tilde{H}\right]=0
\end{equation}
\begin{equation}
\Delta E^{2}=\dfrac{1}{2}\left(\textrm{Tr}\left[\tilde{H}^{2}\right]-\textrm{Tr}\left[\tilde{H}\right]^{2}\right)
\end{equation}
Now, even though our Hamiltonian matrix is truly time dependent, we
must have this equation in particular obeyed during motion, initially
and at the terminal point. Our system is specifically designed using
the linear expansion over the group multipliers such that $\textrm{Tr}\left[\tilde{H}^{2}/2\right]=R^{2}=\sum_{0}^{n}\left|\varepsilon_{j}(t)\right|^{2}$.
As the members of the group are linearly independent and traceless
generators, we can exploit this in the equation above to simplify
to the particular case where
\begin{equation}
\tilde{H}^{3}-R^{2}\tilde{H}-\mathbf{1}\det\left[\tilde{H}\right]=0
\end{equation}
which we can rewrite as $\tilde{H}(\tilde{H}-R\mathbf{1})(\tilde{H}+R\mathbf{1})=\mathbf{1}\det\left[\tilde{H}\right]$.
So we see that the entire behaviour of the physical systems we are
considering depends entirely on the determinant of the Hamiltonian.
Let us exhibit an example of a system with non-zero determinant. We
might have the following:
\begin{equation}
\tilde{H}=\left[\begin{array}{ccc}
\omega_{1} & \varepsilon_{1} & \varepsilon_{2}\\
\overline{\varepsilon}_{1} & \omega_{2} & 0\\
\overline{\varepsilon}_{2} & 0 & \omega_{3}
\end{array}\right]
\end{equation}
In this case, we have $\det\tilde{H}=\omega_{1}\omega_{2}\omega_{3}-\omega_{3}\left|\varepsilon_{1}\right|^{2}-\omega_{2}\left|\varepsilon_{2}\right|^{2}$.
So this system will have different modes of behaviour depending on
the sign of the determinant of the Hamiltonian, which depends on the
difference between the product of the diagonal entries and the weighted
intensities of the control fields. By making $\omega_{1}$ zero we
can actually vary the sign of the determinant by changing the sign
of $\omega_{2}$ and $\omega_{3}$ directly. We notice here the parallels
we are drawing between the energy eigenstate representation, which
does not vary with time, and the periodic behaviour of our previously
examined quantum systems. Indeed, we are able to easily draw results
well known to both mathematicians and physicists alike. We can now
say that the periodicity is shared by the characteristic polynomial,
which is solved at all times by the optimal Hamiltonian and constraint
system. Consider a set of all possible combinations of Hamiltonian
and constraint, for a particular dimension of matrices. We may then
classify all problems into equivalence classes, whose characteristic
polynomials are identical up to permutation of indices over the variables
from the Hamiltonian and constraint. By resolving this at the initial
time, we are able to state confidently that the state will remain
within an equivalence class throughout its periodic evolution. There
may be higher order symmetries that involve cyclic permutation of
the base polynomials between classes in a periodic fashion but to
date this has neither been calculated nor considered and has not arisen
in our calculations thus far.

\section{SU(3) Collective Motion Within $\mathbb{R}^{4}$}

It is worth outlining another related problem that we have solved
in computing the unitary operators and understanding SU(3) in such
fine detail. The question is one of an SU(3) particle embedded in
four dimensions. We refer the reader to recently published works \cite{Morrison 4}
which have examined the nature of time optimal quantum control problems
on SU(4), in particularly the results indicate that it is possible
construct an analogous system of time dependent transformation in
order to reach a Lorentz invariant form of the Dirac equation using
a Hermitian matrix of periodic functions. Other unpublished results
also indicate that the four dimensional angular momentum is brachistochronic;
its matrix time dependence takes the form of a conserved quantity
$\left[\tilde{H},\tilde{M}\right]=0$. This will feature in a future
paper. For now, let us see what it is possible to say about an SU(3)
particle, and how it would behave dynamically as a subspace on four
dimensions. In this situation we have a Hamiltonian matrix that is
of the form:

\begin{equation}
\tilde{H}=\left[\begin{array}{cccc}
0 & 0 & 0 & 0\\
0 & 0 & \varepsilon_{1} & 0\\
0 & \overline{\varepsilon}_{1} & 0 & \varepsilon_{1}\\
0 & 0 & \overline{\varepsilon}_{2} & 0
\end{array}\right]=\left[\begin{array}{cc}
0 & 0\\
0 & \tilde{H}_{3}(t)
\end{array}\right]
\end{equation}
and we will have a constraint matrix that may be written as:
\begin{equation}
\tilde{F}=\left[\begin{array}{cccc}
\omega_{1} & \eta_{1} & \eta_{2} & \eta_{3}\\
\overline{\eta}_{1} & \omega_{2} & 0 & \kappa\\
\overline{\eta}_{2} & 0 & \omega_{3} & 0\\
\overline{\eta}_{3} & \overline{\kappa} & 0 & \omega_{4}
\end{array}\right]
\end{equation}
Computing the quantum brachistochrone $i\dfrac{d}{dt}\left(\tilde{H}+\tilde{F}\right)=\tilde{H}\tilde{F}-\tilde{F}\tilde{H}$
as per usual we find the matrix equations:
\begin{equation}
i\dfrac{d\omega_{j}}{dt}=i\dfrac{d\kappa}{dt}=i\dfrac{d\overline{\kappa}}{dt}=0
\end{equation}
\begin{equation}
i\dfrac{d}{dt}\left[\begin{array}{c}
\eta_{1}\\
\eta_{2}\\
\eta_{3}
\end{array}\right]=\left[\begin{array}{ccc}
0 & \overline{\varepsilon}_{1} & 0\\
\varepsilon_{1} & 0 & \overline{\varepsilon}_{2}\\
0 & \varepsilon_{2} & 0
\end{array}\right]\left[\begin{array}{c}
\eta_{1}\\
\eta_{2}\\
\eta_{3}
\end{array}\right]
\end{equation}
\begin{equation}
i\dfrac{d}{dt}\left[\begin{array}{c}
\varepsilon_{1}\\
\overline{\varepsilon}_{1}\\
\varepsilon_{2}\\
\overline{\varepsilon}_{2}
\end{array}\right]=\left[\begin{array}{cccc}
\omega_{+} & 0 & 0 & -\kappa\\
0 & -\omega_{+} & \overline{\kappa} & 0\\
0 & \kappa & \omega_{-} & 0\\
-\overline{\kappa} & 0 & 0 & -\omega_{-}
\end{array}\right]\left[\begin{array}{c}
\varepsilon_{1}\\
\overline{\varepsilon}_{1}\\
\varepsilon_{2}\\
\overline{\varepsilon}_{2}
\end{array}\right]
\end{equation}
where $\omega_{+}=\omega_{3}-\omega_{2}$, $\omega_{-}=\omega_{4}-\omega_{3}$.
It is clear we are back to playing the same routine as before. We
will be able to calculate time dependence of the Hamiltonian. From
there, we can calculate the time dependence of the state, as it is
a subrotation within SU(4). In this case, the implied scenario is
one of an SU(3) particle evolving within the system Hamiltonian, and
this causes an evolution of an anti-particle in the constraint. SU(3)
violates the compactness property. It is impossible to describe this
space in the same way that the Dirac subalgebra on SU(4) behaves.
We must incorporate an extra transformation to describe the rotation
forwards or backwards in time to describe these particles.

\section{The Failure of Expansion}

We must ask ourselves at this point why people have had so much trouble
before. Although it has been hard work, we have had great success.
However, many of these operators did not exist before we created them
to solve this task. Why was it so difficult? Why have people not investigated
this area before?

The answer lies in the renormalisation theory and quantum electrodynamics,
and the inappropriate use of expansions that work in one area to another
where their domain of application falls short of the required needs.
Let us show exactly how and why this is possible. To begin with, a
central core principal within renormalisation theory is the matrix
expansion:

\begin{equation}
(\hat{A}+\hat{B})^{-1}=\hat{A}^{-1}-\hat{A}^{-1}\hat{B}\hat{A}^{-1}+\hat{A}^{-1}\hat{B}\hat{A}^{-1}\hat{B}\hat{A}^{-1}-...
\end{equation}
used originally in the context of quantum electrodynamics by Feynman
\cite{Feynman 1}. In the case of quantum electrodynamics, we can
write the equation of state for an electron in the form: 
\begin{equation}
\left|\psi\right\rangle =i\hat{H}^{-1}\dfrac{d\left|\psi\right\rangle }{dt}
\end{equation}
and we also have the Hamiltonian, in this case remembering that it
is a constant operator as well as this is time independent quantum
mechanics, given by the matrix operator $\hat{H}=m\hat{\beta}+i\mathbf{p}\cdot\hat{\gamma}$
\cite{Feynman 1,Morrison 4}. Using the expansion above we then can
write the inverse of the Hamiltonian in the form: 
\begin{equation}
\hat{H}^{-1}=\dfrac{\hat{\beta}}{m}-\dfrac{\hat{\beta}}{m}(i\mathbf{p}\cdot\hat{\gamma})\dfrac{\hat{\beta}}{m}+\dfrac{\hat{\beta}}{m}(i\mathbf{p}\cdot\hat{\gamma})\dfrac{\hat{\beta}}{m}(i\mathbf{p}\cdot\hat{\gamma})\dfrac{\hat{\beta}}{m}-...
\end{equation}
To second order we can write out this in terms of initial and final
states as:
\begin{equation}
\left\langle \psi(t)\right|\hat{H}^{-1}\left|\psi(0)\right\rangle \approx T^{(1)}+T^{(2)}
\end{equation}
\begin{equation}
T^{(1)}=\left\langle \psi(t)\right|\dfrac{\hat{\beta}}{m}\left|\psi(0)\right\rangle 
\end{equation}
\begin{equation}
T^{(2)}=-\sum_{\begin{array}{c}
j',k'\\
j'\neq k'
\end{array}}\left\langle \psi(t)\right|\dfrac{\hat{\beta}}{m}\left|j'\right\rangle \left\langle j'\right|i\mathbf{p}\cdot\hat{\gamma}\left|k'\right\rangle \left\langle k'\right|\dfrac{\hat{\beta}}{m}\left|\psi(0)\right\rangle 
\end{equation}
So far this is fine, everything is as normal. One then goes on the
calculate a Lagrangian, shifts in energy values and the like. However,
this entire structure is underpinned by the assumption that $\hat{\beta}^{2}=\mathbf{1}$,
and also that $\hat{H}^{2}=(m^{2}+\left|\mathbf{p}\right|^{2})\mathbf{1}$.
Now, for our SU(3) Hamiltonian matrix, we have something of the form:
\begin{equation}
\hat{H}=\left[\begin{array}{ccc}
0 & \varepsilon_{1} & 0\\
\overline{\varepsilon}_{1} & 0 & \varepsilon_{2}\\
0 & \overline{\varepsilon}_{2} & 0
\end{array}\right]
\end{equation}
where for now we drop the explicit time dependence for convenience.
In this case, we do not have a diagonal element of the form $m\hat{\beta}$
forming part of the Hamiltonian which drives the dynamics of state,
and even worse we have an element: 
\begin{equation}
\hat{H}^{2}=\left[\begin{array}{ccc}
\left|\varepsilon_{1}\right|^{2} & 0 & \varepsilon_{1}\varepsilon_{2}\\
0 & \left|\varepsilon_{1}\right|^{2}+\left|\varepsilon_{2}\right|^{2} & 0\\
\overline{\varepsilon}_{1}\overline{\varepsilon}_{2} & 0 & \left|\varepsilon_{2}\right|^{2}
\end{array}\right]
\end{equation}
which is immediately more difficult to handle. In attempt to remedy
the situation, various attempts have been made to salvage the expansion,
which since the $\hat{H}$ for SU(3) does not have a diagonal component,
one is manually added in, and subtracted from the other component.
One can see how this calculation immediately blows up. What is missing
here is the understanding of how the constraint and the system are
interacting wholistically. We can see immediately from the above formula
that we will have components of $\hat{H}^{2}$ interacting with the
constraint $\hat{F}$. By only considering the dynamic evolution from
the Hamiltonian, we are not understanding the physics properly. That
it works in a certain instance is due to a pecularity of SU(4) and
its special block factorisation group. That method is special and
unique to that group, and may work perfectly for other groups that
have similar properties. It will not suffice for the sort of complex
group that SU(3) describes.

\section{Discussion}

We have shown in this paper how effective the use of matrix calculus
may be when developed in order to understand SU(3). This has enabled
us to re-examine the nature of the quantum state within this environment.
The answers are intriguing. With simple precepts we have been able
to develop an understanding of how transformation laws act on the
fundamental states which describe the space. These transformation
laws then allow the derivation of the relationships between fundamental
particles. A natural transformation that emerges on this set, which
we have been able to understand in light of the time-reversal asymmetry
for this space.

We must talk a little more about the operator which transforms between
the Floquet picture, where everything rotates around a static particle,
to the picture where the particle moves and everything else stays
still. This operator, which we have written throughout as the matrix
$\hat{Y}$, represents a fundamental physical symmetry which we can
associate to the nuclear interactions. It tells us that there will
be another quantum number by which we can classify the relationships
between the groups of particles that span the space. It also implies
that there will be certain results in terms of parity transformations,
which we have known for some time to be an experimentally verified
fact post the original experiments of Wu \cite{Wu}, and theoretical
discussion of Yang and Lee \cite{Lee} . That such a fundamental parameter
which has been a matter of intrigue arises naturally from our calculation
method is a pleasing surprise. If it did not, we would have reason
to invalidate our theory. That we are not justified in doing so, shall
be the result of experiment. These types of exotic states can be expected
to occur in high energy scattering experiments. In particular, we
have solved the mysteries of the standard model. It arises naturally
as the dynamics of the generating group of this complex projective
space as it evolves in time through a periodic motion. The implied
symmetry in this space takes the form of being either from the outside,
looking in; or from the inside looking out. The transformation laws
then imply that the time evolution operator must incorporate this
extra transformation $\hat{Y}$, in order to recover the correct matrices
for forward-in-time and backward-in-time motion. That the direction
of time implies a certain helicity of the state is known; however,
it is uncommon to see it expressed in such a fundamental way. The
matrix calculus developed has been of no end of aid to this cause.

The problem that has been addressed in this calculation is more general
than it would appear on face value. Indeed, it is possible to show
that the results given will apply to any four dimensional subspace
of SU(3), by virtue of the permutation of root systems. The example
we have considered is most general, and all non-trivial subspaces
of SU(3) are either permutations of the problem considered, or projections
to a lower order space within the greater group. In this sense, we
can illustrate most clearly the difference between the dynamics of
these types of states and those on even-numbered dimensions. Because
we are forced to deal with the degenerate element, and the implied
centrality of the system, the time evolution operator is not solely
a reflection of the time translation of states. In fact, we must deal
properly with the plethora of transformations that are generated by
this group, and find the one element which distinguishes particles
from one another. 

This example of time optimal quantum control has been difficult to
treat. The essential degeneracy and multi-valued nature of the underlying
group has required a number of tricky transformation methods to be
used in order to reach resolution. While to expect that a complication
of a scenario would be unlikely to achieve better results, through
this calculation we have gained a better understanding of the nature
of the dynamical laws that govern these types of systems. Indeed,
it is simple to see how similar results will apply to any quantum
system of odd dimension. Any system of odd matrix dimension will have
a degenerate centralising element, and the physics should be worked
out in a similar way to how we have proceeded in this paper. Whether
that is the case in reality, we shall see. 

Our understanding of the nature of symmetry is thus expanded; from
this simple example we have constructed a complex insight into the
nature of fundamental particles. We must comment on the fundamental
difference here. This space, as it is no longer simply connected,
has a certain geometrical complexity that arises within the time evolution
operator. Due to this peculiarity of the geometrical arrangement,
we see that decomposition of the Hamiltonian matrix into the diagonal
representation results in the time evolution operator separating into
two equivalent rotations common to the unitary and its adjoint, and
an isometric transformation depending on whether one is going forward
or backward in time. This is the fundamental and defining feature
of this space and is not observed on some other physically relevant
quantum control problems involving relativistic electronic states. 

Now, since it all seems as if we started out on an investigation of
some dynamic methods, and all of a sudden are talking about subatomic
particles, let us now focus on some potential applications of this
theory. We can actually exploit this property of multiple degeneracy
to achieve better results in a functional quantum computer, or at
least for better forms of quantum control. We have shown how all the
gates can be constructed, and all the useful things that can be done.
Even stranger than this, the notion of a zero state could be used
in the following way. We could construct an operator $\hat{A}$ such
that we know $\hat{A}\left|0\right\rangle =0$. It might even be that
$\hat{A}(t)\left|0\right\rangle =\lambda(t)$ and we know that at
particular times with high probability, the function $\lambda(t)$
is zero. We can use this to effectively void quantum states which
are involved in our calculation, dumping intermediate steps and bad
data alike into the $\left|0\right\rangle $ state. We then can- at
the conclusion of the calculation- just project onto another state,
which we know for certain is orthogonal to the degenerate state. This
will enable the operation of an effective fan-out step for quantum
computation. We know this state has an effective energy of zero, at
least to first approximation. 

The focus of a large body of work in the quantum computation domain
is solely devoted to the use and implementation of quantum logic in
two state systems. This example has highlighted the difference in
behaviour for even small modifications of the precepts related to
quantum computing. It has been a fruitful exercise. We have learned
of degenerate eigenspaces, how unitary transformations interlace with
isometries, the strange particles and how their transformation laws
relate to the geometry of the system and constraint within SU(3).
Other ways in which this unique geometry could be exploited to gain
are obvious. The unitary transforms and the contact transformations
are immediately comparable to other quantum gates such as the Hadamard
matrix, with the added bonus of an ancilla. This can be seen as a
rotation within the SU(2) subgroup which is part of our greater system.

This paper may have seemed laborious at times due to the non-commutative
nature of the matrix multiplication. In that regard, the aid of computers
and symbolic algebra has been of indisputable aid in avoiding the
almost insufferable arithmetic. Readers who are interested in the
implementation of constructions used in this paper using symbolic
algebra are invited to contact the author for further discussion.
One aspect of these types of systems with geometric symmetry that
is of most use and guidance when carrying out long calculations is
the nature of error correction. The symmetry acts in a way to correct
the calculation back towards its proper centre, with the phase acting
as a trace or dye. The collapse of proper phase and order within the
matrices themselves is then an indication that the calculation is
awry.

Feynman states in his notes that ``...for the new strange particles,
we have no idea what $H_{ij}$'s to use. In other words, no one knows
the \textit{complete} $H_{ij}$ for the whole world. (P)art of the
difficulty is that one can hardly hope to discover the $H_{ij}$ when
no one even knows what the base states are!'' \cite{Feynman 2}.
With this calculation, we have moved one step closer towards achieving
that objective. We have demonstrated a degree of complete quantum
control, if only at particular instants of time. We have managed to
extract the fundamental symmetries of this system using a simple dynamic
consideration that may be applied to many other different situations,
each with their own degree of complexity and inherent fundamental
properties and conservation laws. Furthermore, he then extends that
claim to propose that we can imagine one equation being converted
into another if ``...we replace the classical energy by the Hamiltonian
and the classical $\mu$ by the matrix $\mu\sigma$. Then, after this
purely formal substitution, we interpret the result as a matrix equation''.
He then emphasises that ``..it is really more correct to say that
the Hamiltonian matrix corresponds to the energy, and any quantity
that can be defined with a corresponding matrix'' \cite{Feynman 2}.
We must agree with the former, and not the latter. We have shown with
our methods and calculations that there is no longer any need to guess
the Hamiltonian matrix which describes the energy. All we need is
the constraints that describe the quantum system in order to capture
the behaviour.

\section{Further Directions}

\subsection{Experimental tests}

This is by far the most important place where gains will be made now
that we have an understanding of the mechanics of the internal states
of these particles. The results presented both here and within should
allow a thorough and complete investigation to be made of the dynamics
of the nuclear states, particle decay etc. This will not only be in
the form of high-energy particle experiments. We have already discussed
the triple quantum dot; we shall not go over that again. Other experiments
should look at the links between computer simulated experiments using
continuous Hamiltonian operators and expanding these methodologies
to match the computer experiments. One immediate place this could
be achieved is in lattice-QCD scenarios. This will enable good tests
and comparisons to be made both between the predictions from this
theory, the breakdowns in computation, and the middle ground that
will contain regions of validity. 

One crucial experimental test that can be carried out is the determination
of the parameter $\theta$. This parameter describes the relative
asymmetry these subatomic particles experience when travelling forwards
and backwards in time. Once this has been cleared up, the measurement
of all other parameters for the atomic states can be placed within
context. There may be some consolidation of data in this process.

Another way in which this time asymmetry could be examined is examining
high energy particle tracks observed in cloud chambers. One could
then simply treat it as if it was a particle moving backwards in time
with appropriate isometry by reversing the trajectory. This might
make an effective comparison for the time asymmetry parameter to be
determined statistically.

There is the distinct possibility that a major theory may be proven
false through empirical trial. For this reason, this task must take
first priority.

\subsection{Relating to a certain particle}

Interesting problems still remain on the $5\times5$ matrices for
a certain spin-2 particle. We can expect to find an eigenvalue problem
which will be written as $\tilde{H}(\tilde{H}^{2}-R_{1}^{2})(\tilde{H}^{2}-R_{2}^{2})=0$.
This problem will contain an intertwining operator concatenated with
some form of $\hat{Y}$ peculiar to the space, overlaying some form
of the two operators $\hat{U}_{1}$ and $\hat{U}_{2}$. One immediately
sees that this will be a nested version of all the problems we have
already engaged on SU(2), SU(3) and SU(4) while having its own wrinkles
that make it unique in and of itself. This will be much more complicated
due to size, and the level of twisting that will be required to understand
and decouple its equations of motion. Finding the space of states
will be complicated further by the fact that we do not know what the
symmetries of the interaction are. Some preliminary work has already
been done, and appears promising. The results will have implications
for the science of gravity if successful. The lack of experimental
evidence will make this task more difficult. In calculating these
examples for the electromagnetic, strong and weak nuclear forces,
we have had the aid of a large amount of data, knowledge and empirical
understanding. However, the known constants for gravity and what its
particular symmetries are other than the inverse square law, remain
shrouded due to the weakness of the interaction. The principal axis
of attack to address this problem will be to find a way in which to
describe the maximal physics of the interaction, while using a minimum
of operators via the constraint law.

\subsection{Metric tensors}

These calculations have produced a wealth of unitary operators, but
there are significant areas that need to be re-examined now that we
have made the initial pass. This whole method is underpinned by the
Fubini-Study metric \cite{Carlini 1,Carlini 2}, which has the following
property: it is a valid metric tensor that satisfies the equations
of general relativity. Now that we have a lock on the method used
to produce the Hamiltonian, we should look closely at what metrics
are implied by these sort of quantum geodesics and see whether there
are broader implications for the art of general relativity.

\subsection{Continuum mechanics}

Let us now move to the consideration of an important task that we
have not analysed in detail in this paper. Indeed, we have completely
neglected its existence, which is crucial to the nature of scattered
states and other ways in which this formalism can be extended. We
must talk of the topic of time independent quantum mechanics, and
in particular, the nature of a continuous variable. Given the results
presented, we must re-examine the premises and foundations upon which
we have been making our gauge of reality, and whether continous variables
can be considered to exist. However, let us put these qualms to one
side and merely discuss what might be done about the situation. Instead
of trying to fight one system against another, consider the following
middle ground. The continuous analog of the quantum brachistochrone
is not well defined. Let us show how to set up the problem so that
it can be understood in a consistent fashion with what we have just
achieved. We have the quantum brachistochrone viz:
\begin{equation}
i\dfrac{d}{dt}\left(\tilde{H}+\tilde{F}\right)=\tilde{H}\tilde{F}-\tilde{F}\tilde{H}
\end{equation}
as well as the Schrödinger equation for the state: 
\begin{equation}
i\dfrac{d\left|\Psi\right\rangle }{dt}=\tilde{H}(t)\left|\Psi(t)\right\rangle 
\end{equation}

To go about constructing a viable form of the quantum brachistochrone
suitable for continuum quantum mechanics, we shall do as follows.
We must be able to define the equations of motion through some variant
of the Schrödinger equation as shown above. Next, as we are not dealing
with a situation of time changing, the underlying parameter which
we vary against must be the variable in the equation that replaces
time. Third, we must have constraints represented in a continuous
fashion, that describe the underlying state and the behaviour of the
potential that defines it. We expect that the conservation of energy,
given by the Schrödinger equation, should emerge naturally from the
calculation of the quantum brachistochrone.

Given that in the original quantum control problem, we have a Hamiltonian
and a constraint, it is reasonable to ask what choice and role will
be played by the constraint. Constraints for a continuous state are
more complicated, but can be seen this way. A constraint is a boundary
condition for an integral of a sort; its nature excludes the state
from a zone of influence. A constraint might be a single point at
infinity, or minus infinity, which we know that the state never visits.
This causes the Taylor series which defines its expansion to converge.
By a result similar to that of \cite{Novikov}, we will have a reasonable
series if the Plancheral identity closes and is less than infinity.
This will be able to occur as long as the state has either a deleted
half-plane, two deleted halves either side of an allowed zone, or
a combination of either one or two deleted points at infinity with
a deleted half plane. So the conditions for the constraint must read
something like:
\begin{equation}
\int_{S}\psi^{\star}(x)\psi(x)dx=0
\end{equation}
or 
\begin{equation}
\lim_{x\rightarrow a}V(x)=\infty
\end{equation}
where $S$ is a region of integration where the state is excluded
from by the potential. Continuing this exercise, we must have the
operator $i\dfrac{d}{dt}$ replaced by $i\dfrac{\partial}{\partial q}$.
Now, the Hamiltonian in the time-dependent quantum control problem
is the matrix operator which takes the state from one time to another.
However, we are now in a situation where the situation is not changing
with respect to time. It is now changing with respect to the parameter
space. So this choice is justified. As we can see by the argument
above, the constraint is nothing more than the potential which confines
the state. We can vary its strength and dependence on the underlying
parameter space and alter the behaviour and consistency of the quantum
state. So the substitution of the matrix operator $\hat{F}$ will
be by its continuum counterpart, the continuous potential $\hat{V}(p,q)$.
The next part of the calculation is the isotropic condition; in that
case we require an operator which drives the dynamics to have the
property $\textrm{Tr}\left[\tilde{H}^{2}\right]/2=k$. However, continuing
with the analogy, we must therefore have another condition on the
operator which drives the dynamics. In the continuous case, this will
take the case of a Plancheral identity related to the momentum via:
\begin{equation}
\int_{S}\psi^{\star}(q)\hat{\epsilon}^{2}\psi(q)dx=k<\infty
\end{equation}
We have a quantum brachistochrone equation already. Let's put the
pieces together. For the finite system, we have the equation: 
\begin{equation}
i\dfrac{d}{dt}\left(\tilde{H}+\tilde{F}\right)=\tilde{H}\tilde{F}-\tilde{F}\tilde{H}
\end{equation}
and the constraints 
\begin{equation}
\left\{ \begin{array}{c}
\textrm{Tr}\left[\tilde{H}\tilde{F}\right]=0\\
\\
\textrm{Tr}\left[\tilde{H}^{2}\right]/2=R^{2}<\infty
\end{array}\right.
\end{equation}

Now, a quick note. In this system, we have dynamics that are described
via a Schrödinger equation related to the momentum. We would have
the following momentum evolution equation, which describes the momentum
as it evolves in space:
\begin{equation}
\hat{p}\psi=\hat{\epsilon}(p,q)\psi
\end{equation}

\begin{equation}
i\dfrac{\partial\psi}{\partial q}=\hat{p}\psi
\end{equation}
where $\hat{\epsilon}(p,q)$ are essentially the Lie group coefficients
which give the momentum operators for the group as translations of
the fundamental differential operators. We view $\hat{\epsilon}(p,q)$
as an infinite dimensional square matrix which represents a function,
and $\psi$ as a normalisable infinite dimensional column vector,
which means that the coefficients die off sufficiently fast as the
function travels to infinity. Replacing the pieces using the prescription
above, we would obtain the following:
\begin{equation}
i\dfrac{\partial}{\partial q}\left(\hat{p}+\hat{V}\right)=\hat{p}\hat{V}-\hat{V}\hat{p}
\end{equation}
and a constraint system that would read as:
\begin{equation}
\left\{ \begin{array}{c}
\int_{S}\psi^{*}\hat{V}(q)\hat{p}(p,q)\psi dq=0\\
\\
\dfrac{1}{2}\int_{S}\psi^{*}\hat{\epsilon}^{2}\psi dq=R^{2}<\infty
\end{array}\right.
\end{equation}
Now, this is not quite correct. It is almost correct, but as it stands,
technically eq. (128) should be rewritten as the expression below:
\begin{equation}
\begin{array}{c}
\int_{S}\psi^{\star}(q)\left[i\dfrac{\partial}{\partial q}\left(\hat{p}+\hat{V}\right)\right]\psi(q)dq\\
\\
=\int_{S}\psi^{\star}(q)\left[\hat{p}\hat{V}-\hat{V}\hat{p}\right]\psi(q)dq
\end{array}
\end{equation}

We must remember that in this representation of quantum mechanics,
the objects that we have been using as matrices go over into operators
of functions and differential operators, and the commutation rules
are generated through the composition of the functions. With this
in mind, we can easily see what is going on here. For the first of
the constraints, and considering we have a sufficiently smooth function,
we will be able to replace this by the expression $\int_{S}\psi^{*}\hat{V}(x)\psi dx=0$
which we can easily evaluate as a region where the potential forbids
the states as discussed before. The second condition holds that the
variance in the total momentum is held to some absolute maximum less
than infinity, as with our other problem. This intuitively feels the
same in nature to the quantum brachistochrone equation for finite
systems. Nothing changes in the Fubini-Study metric, as we have the
Plancheral identity, everything is still fine. However, the uncertainty
that was modelled in the time-dependent system is now the uncertainty
in the momentum of the state, so we will have an identity like $\Delta p_{q}=\dfrac{ds}{dq}$
which we can use to get the metric for the continuous state space
via $\int1dq+constraints=\min$ as before.

So now we are in the strange and unsettling place where we see the
full consequences of this question of quantum control of atomic states.
We find that the harder we try to control it, the more it expands
in uncertainty. Indeed, the potential and the state are indistinguishable,
by this relationship. This intuitively fits with the ideas that this
paper has explored, as well as another recent work. It has proven
impossible to separate the system from the environment. 

We have succeeded in a complicated task. This calculation has shown
how we can nest subsystems within a greater geometry and come up with
testable dynamic conclusions. We have also conclusively illustrated
some strange facts in full. For example, we can see that from one
perspective, the constraint is a constant applied at each time with
a matrix variable that drives the energy evolving through it. This
implies a certain dynamical symmetry on the Hamiltonian matrix; the
consistency of these two viewpoints is the crux of the physics and
the key to understanding this behaviour. On the one hand, we have
a matrix which is driving the energy, and a set of forbidden states,
which we call the constraint. The matrix is changing in time, the
constraint staying fixed. On the other hand, we could view the system
as imparting energy to the set of forbidden states at some initial
point and driving the evolution of the constraint through time. Reconciliation
of these two seemingly disparate perspectives produces all the matrix
calculus we have presented in this paper.

\bibliographystyle{alpha}  \bibliographystyle{alpha}
\bibliography{abbrv,textbooks,QuantumComputation,stallisEntropy}

\appendix

\section{Transformations \& Matrices}

\subsection{Angular momentum matrices}

\begin{equation}
\hat{L}_{x}=\left[\begin{array}{ccc}
0 & 1 & 0\\
1 & 0 & 1\\
0 & 1 & 0
\end{array}\right]
\end{equation}
\begin{equation}
\hat{L}_{y}=\left[\begin{array}{ccc}
0 & -i & 0\\
i & 0 & -i\\
0 & i & 0
\end{array}\right]
\end{equation}
\begin{equation}
\hat{L}_{z}=\left[\begin{array}{ccc}
1 & 0 & 0\\
0 & 0 & 0\\
0 & 0 & -1
\end{array}\right]
\end{equation}

\subsection{Solution matrices for SU(3)}

\begin{flushleft}
These matrices have a fundamental similarity to the Bogoliubov transform
and can be classified as either fermion-type or boson-type on the
basis of the determinant. Due to the null state the central element
is fixed to be 1.
\begin{equation}
\hat{X}_{Q}(t)=\left[\begin{array}{ccc}
\dfrac{1}{\sqrt{2}}\cos t & -\dfrac{1}{\sqrt{2}}\cos t & ie^{-i\theta}\sin t\\
\dfrac{1}{\sqrt{2}} & \dfrac{1}{\sqrt{2}} & 0\\
\dfrac{i}{\sqrt{2}}e^{i\theta}\sin t & -\dfrac{i}{\sqrt{2}}e^{i\theta}\sin t & \cos t
\end{array}\right]
\end{equation}
 
\begin{equation}
\hat{X}_{J}(t)=\left[\begin{array}{ccc}
\dfrac{1}{\sqrt{2}}\cos t & -\dfrac{1}{\sqrt{2}}\cos t & -\sin t\\
\dfrac{1}{\sqrt{2}} & \dfrac{1}{\sqrt{2}} & 0\\
\dfrac{i}{\sqrt{2}}\sin t & -\dfrac{i}{\sqrt{2}}\sin t & i\cos t
\end{array}\right]
\end{equation}
\begin{equation}
\hat{X}_{D}(t)=\left[\begin{array}{ccc}
-\dfrac{i}{\sqrt{2}}\cos t & \dfrac{i}{\sqrt{2}}\cos t & i\sin t\\
\dfrac{1}{\sqrt{2}} & \dfrac{1}{\sqrt{2}} & 0\\
\dfrac{i}{\sqrt{2}}\sin t & -\dfrac{i}{\sqrt{2}}\sin t & i\cos t
\end{array}\right]
\end{equation}

\par\end{flushleft}

\appendix

\subsection{Unitary Transformations}

We list the transformations used in Section VII. For the solution
matrix $\hat{X}_{Q}$:

\begin{flushleft}
\begin{equation}
\hat{R}_{q_{1}}(\sigma)=\left[\begin{array}{ccc}
\cos\sigma & 0 & -ie^{-i\theta}\sin\sigma\\
0 & 1 & 0\\
+ie^{i\theta}\sin\sigma & 0 & \cos\sigma
\end{array}\right]
\end{equation}
\begin{equation}
\hat{R}_{q_{2}}(\sigma)=\left[\begin{array}{ccc}
\cos\sigma & 0 & -ie^{-i\theta}\sin\sigma\\
0 & 1 & 0\\
-ie^{i\theta}\sin\sigma & 0 & \cos\sigma
\end{array}\right]
\end{equation}
\begin{equation}
\hat{R}_{q_{3}}(\sigma)=\left[\begin{array}{ccc}
\cos\sigma & 0 & ie^{-i\theta}\sin\sigma\\
0 & 1 & 0\\
-ie^{i\theta}\sin\sigma & 0 & \cos\sigma
\end{array}\right]=\hat{R}_{q_{1}}(-\sigma)
\end{equation}
For the solution matrix $\hat{X}_{J}$:
\begin{equation}
\hat{R}_{j_{1}}(\sigma)=\left[\begin{array}{ccc}
\cos\sigma & 0 & i\sin\sigma\\
0 & 1 & 0\\
i\sin\sigma & 0 & \cos\sigma
\end{array}\right]
\end{equation}
\begin{equation}
\hat{R}_{j_{2}}(\sigma)=\left[\begin{array}{ccc}
\cos\sigma & 0 & -i\sin\sigma\\
0 & 1 & 0\\
i\sin\sigma & 0 & \cos\sigma
\end{array}\right]
\end{equation}
\begin{equation}
\hat{R}_{j_{3}}(\sigma)=\hat{R}_{j_{2}}(\sigma)
\end{equation}
To find a set of rotations for this subspace, note that $\hat{R}_{j_{2}}(-\sigma)$
is distinguished from the other members of the group. We may therefore
form the set $\left\{ \hat{R}_{j_{1}}(\sigma),\hat{R}_{j_{2}}(\sigma),\hat{R}_{j_{2}}(-\sigma)\right\} $
to describe the rotations of these states within the subspace. For
the solution matrix $\hat{X}_{D}$:
\begin{equation}
\hat{R}_{d_{1}}(\sigma)=\left[\begin{array}{ccc}
\cos\sigma & 0 & \sin\sigma\\
0 & 1 & 0\\
-\sin\sigma & 0 & \cos\sigma
\end{array}\right]
\end{equation}
\begin{equation}
\hat{R}_{d_{3}}(\sigma)=\hat{R}_{d_{2}}(\sigma)=\hat{R}_{d_{1}}(\sigma)
\end{equation}
This space can be described by the set of rotations as$\left\{ \hat{R}_{d_{1}}(\sigma),\hat{R}_{d_{1}}(-\sigma)\right\} $.
The set of 8 rotations plus the unitary operator forms a set for the
group. From this it is possible to write down all the transformation
laws against the fundamental unitary operator, obviously only the
unitary operator involved in both the forward in time and backwards
in time evolution operator will transform as itself.
\par\end{flushleft}
\end{document}